\documentclass[iop,apj,numberedappendix,appendixfloats]{emulateapj}
\usepackage{amssymb,amsmath,amsthm}
\usepackage{amsmath, xspace}
\usepackage{graphics,graphicx} 
\usepackage{rotating}
\usepackage{color}
\usepackage{perpage}
\begin{document}

\shortauthors{Zheng et al.}
\shorttitle{Radio Sources in the NCP Region}

\slugcomment{Received xxx; accepted xxx; published 2016 xxx}

\title{Radio Sources in the NCP Region Observed with the 21 Centimeter Array}

\author{
Qian Zheng\altaffilmark{1,2,3},
Xiang-Ping Wu\altaffilmark{2,3,4},
Melanie Johnston-Hollitt\altaffilmark{1},
Jun-hua Gu\altaffilmark{2,3},
and
Haiguang Xu\altaffilmark{5}
}

\email{wxp@bao.ac.cn}
\altaffiltext{1}{School of Chemical and Physical Sciences, PO Box 600, 
Victoria University of Wellington, Wellington 6140, New Zealand\\}
\altaffiltext{2}{National Astronomical Observatories, Chinese Academy of 
Sciences, 20A Datun Road, Beijing 100012, China\\}
\altaffiltext{3}{Center for Astronomical Mega-Science, Chinese Academy of 
Sciences, 20A Datun Road, Beijing 100012, China\\}
\altaffiltext{4}{Shanghai Astronomical Observatory, Chinese Academy of 
Sciences, 80 Nandan Road, Shanghai 200030, China\\}
\altaffiltext{5}{Department of Physics and Astronomy, Shanghai Jiao Tong 
University, 800 Dongchuan Road, Minhang, Shanghai,200240,China\\}


\begin{abstract}

We present a catalog of 624 radio sources detected around the North Celestial 
Pole (NCP) with the 21 Centimeter Array (21CMA), a radio interferometer 
dedicated to the statistical measurement of the epoch of reionization (EoR). 
The data are taken from a 12 h observation made on 2013 April 13,  
with a frequency coverage from 75 to 175 MHz and an angular resolution
of $\sim4^{\prime}$. The catalog includes flux densities at eight sub-bands
across the 21CMA bandwidth and provides the in-band spectral indices
for the detected sources. To reduce the complexity of interferometric 
imaging from the so-called ``w'' term and ionospheric effects, the present 
analysis are restricted to the east-west baselines within 1500 m 
only. The 624 radio sources are found within 5 degrees around the NCP 
down to $\sim0.1$ Jy. Our source counts are compared, and also exhibit 
a good agreement, with deep low-frequency observations made recently 
with the GMRT and MWA. In particular, for fainter radio sources 
below $\sim1$ Jy, we find a flattening trend of source counts 
towards lower frequencies. While the thermal noise ($\sim0.4$ mJy) 
is well controlled to below the confusion limit, the dynamical 
range ($\sim10^4$) and sensitivity of current 21CMA imaging is largely 
limited by calibration and deconvolution errors, especially the grating 
lobes of very bright sources, such as 3C061.1, in the NCP field 
which result from
the regular spacings of the 21CMA. We note that particular attention 
should be paid to the extended sources, and their modeling and removals 
may constitute a large technical challenge for current EoR experiments. 
Our analysis may serve as a useful guide to design of next generation 
low-frequency interferometers like the Square Kilometre Array.

\end{abstract}
\keywords{radio continuum: general---instrumentation: 
interferometric---radio continuum: galaxies---methods: observational}


\section{Introduction}
\label{sec:intro}

Low-frequency observations are very important for the study of the 
statistical signatures of extragalactic sources, in particular source 
counts in the low-frequency sky, which until recently have been only 
characterized at the very highest end of the flux scale. The number 
of steep-spectrum sources increases rapidly as the sample selection 
frequency is lowered \citep{Massaro14}, because these sources become 
too faint to be detected at high frequencies leading to an expectation 
that low-frequency sky surveys will uncover this population in greater 
numbers than current cm-wavelength studies. As it is not trivial to infer 
the low-frequency ($\nu \leq$ 300 MHz) sky from higher frequency data, 
deep source counts at the same frequency as EoR observations are required 
for accurate foreground modeling and subtraction. Better understanding 
of the statistical properties and behavior of radio sources is required 
and motivated by a number of science goals, such as the foreground modeling 
and subtraction in the detection of the EoR.

Over the last 30 years, a large and ever increasing number of surveys and 
targeted observations have been performed at low frequencies commencing 
with the 6th Cambridge (6C) Survey of Radio Sources \citep{Hales07}, 
which covered the entire northern sky above 30 degrees at 151 MHz. 
Since then a number of other surveys have been completed such as the 7th 
and 8th Cambridge Survey of Radio Sources (7C, and 8C: \citealt{Rees90}), 
the targeted survey on the Culgoora Circular Array \citep{Slee95} in which 
the flux densities of 1800 high-frequency-selected radio sources were measured 
at 80 and 160 MHz, the 74 MHz VLA Low-Frequency Sky Survey 
(VLSS: \citealt{Cohen07}) above -30 degrees, the Murchison Widefield 
Array Commissioning Survey (MWACS: \citealt{NHW14}) covering 
6,100 $\rm{deg}^2$ between 104 and 196 MHz, and the Precision Array for 
Probing the Epoch of Reionization (PAPER; \citealt{Jacobs11}) low resolution 
survey of the Southern sky at 145 MHz. Current low frequency radio 
interferometers, such as LOFAR \citep{vH13}, GMRT \citep{Paciga13}, 
LWA \citep{Taylor12}, MWA \citep{Bowman13,Tingay13} and PAPER \citep{Jacobs11} 
have made a considerable advance in our understanding of the population of 
extragalactic radio sources at frequencies below 300 MHz with all-sky low 
frequency surveys(\citealt{Wayth15,Heald15,Intema16}). The upcoming Square Kilometre 
Array (SKA) is expected to continue to push the boundaries of low-frequency 
astronomy, observing the radio sky at much higher sensitivity and resolution. 

Recently, \citet{Franzen16} determined the 154MHz source counts with
MWA observation of the so-called EoR0 field (which is centered at J2000 
$\alpha=00^h00^m00^s,\delta=-27^o00'00''$) and compared these with the 7C counts 
at 151MHz \citep{Hales07}, as well as the GMRT source counts at 153MHz 
\citep{Intema11, Ghosh12, Williams13} finding good agreement to a limiting 
flux density of $\sim$40 mJy. Using LOFAR, \citet{vanWeeren14} examined source 
counts at 34, 46 and 62 MHz, a wavelength regime not explored since the 
8C observations at 38 MHz 25 years earlier. The LOFAR observations are the 
deepest images ever obtained at these frequencies and reach noise levels 
of the order of a few mJy, being nearly 2 orders of magnitude deeper than 
the 8C survey. \citet{vanWeeren14} confirmed the previously 
known result that the average spectral index of radio galaxies flattens 
towards lower frequencies, as is expected from synchrotron losses 
\citep{Franzen16}. In addition to understanding the properties of 
the low frequency radio population to study the population itself, 
it is of great importance to properly characterize the population if 
we are to fully exploit these instruments to detect the EoR as a precise 
sky model is vital for subtracting foreground sources 
\citep{Jelic08,Ghosh12,Moore13,Jelic14,Thyagarajan15,Offringa16}. Thus, source 
count studies of the low frequency sky are strongly motivated by a number 
of complementary science goals.

The North Celestial Pole (NCP) region is covered at low-frequencies 
by the WSRT \citep{Bernardi10} and LOFAR \citep{Yatawatta13}, it is 
also the sky region targeted by the 21 Centimeter Array (21CMA). 
The key science goal for the 21CMA is to statistically measure 
the redshifted 21cm signal of neutral hydrogen from the EoR. 
Determining the radio source counts and understanding the spectral 
properties of the radio sources are important for removing the 
foreground sources and extracting the faint EoR signal. In this paper, 
we present targeted observations centered on the NCP and extending 
5 degrees in radius using the 21CMA. We determine the source counts 
using eight sub-band images between 75 MHz and 175 MHz with flux 
densities down to $\sim 0.1$ Jy. We also investigate the spectral properties 
of catalog sources and compare our source counts with the recent 
surveys. 

The outline of the paper is as follows. Section 2 describes the 21CMA 
telescope, the observation and data reduction pipeline. Section 3 presents 
the source extraction process and resultant source catalog and properties 
including number counts, spectral indices, completeness and errors. 
Discussion and concluding remarks are given in Section 4. 

\section{Observations and Data Reduction}
\label{sec:obs}
\subsection{The 21CMA}

\begin{figure}
\begin{center}
\hspace{-4mm}
\includegraphics[width=8cm]{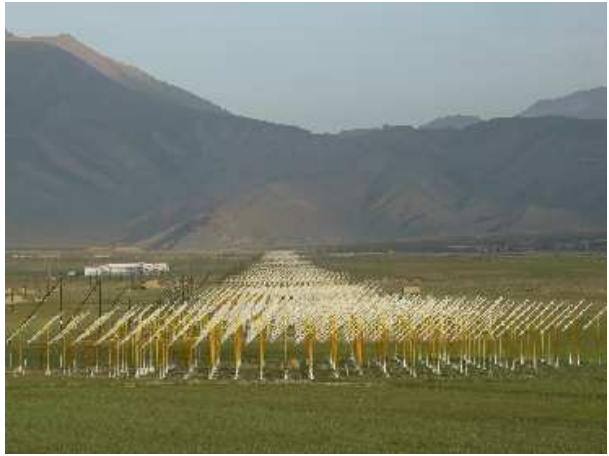}
\vspace{3mm}
\caption{Photograph of the east-west baseline of the 21CMA, in which
the building to the left is the control center.}
\vspace{3mm}
\label{fig:baseline}
\end{center}
\end{figure}

The 21CMA is a ground based radio interferometer dedicated to detection of 
the EoR. The array, sited in the Ulastai valley of western China, consists 
of 81 pods or stations, with a total of 10287 log-periodic antennas are 
deployed in two perpendicular arms along an east-west (6.1 km) 
(Figure \ref{fig:baseline}) and north-south (4 km) direction, respectively. 
Spacing of these 81 pods is chosen such that a sufficiently large number of
redundant baselines and a good uniform uv coverage can both be guaranteed. 
Each antenna element has 16 pairs of dipoles with lengths varying from 
0.242 m to 0.829 m, optimized to cover a frequency range of 50-200 MHz,
which gives rise to an angular resolution of $3'$ at 200 MHz. 
All the antennas are fixed on the ground and point at the NCP for 
the sake of simplicity and economy, which allows us to observe the same patch 
of sky for 24 hours a day through the whole year. Therefore, we can reach 
a higher sensitivity in a region of a few ten square degrees around the 
NCP in a relatively short time. Phase delayed Coaxial cables are used to 
combine the signal from each antenna for each pod, and the output signal is 
digitized at a sampling rate of 400 MHz with 8 bit precision to 
accommodate the bandwidth of 200 MHz, although signal below 50 MHz is 
actually filtered out. Fast Fourier Transforms (FFT) and correlations 
are performed in software by a cluster of 83, dual-core Intel(R) Xeon(TM) 
servers, equipped with gigabit ethernet networks (PCI-E 10-Gbps 4X) and 
a high speed network (Infiniband) switch (CISCO SFS-7008P). The 200 MHz 
bandwidth is divided into 8192 channels, giving rise to a frequency 
resolution of 24.4 kHz. Data in each channel are integrated for about 
3 seconds in memory before they are output to a disk array of 32 terabytes.
Hard disks are transported to the headquarters of the National 
Astronomical Observatories of China in Beijing for analysis.

Construction of the 21CMA was completed in 2006, and a significant 
upgrade was made in 2009 by installing a new type of low noise amplifier 
(40 K) on each antenna and 40 sets of GPUs in a new data acquisition 
cluster. The former considerably suppresses thermal noise from the phased 
delayed cables, while the later provides a very efficient way to speed 
up correlation computation processing. We have used software instead 
of hardware or FPGAs to perform the FFT and correlations. After this 
upgrade we also decided to run the E-W baseline only to maintain 
a higher efficiency ($50\%$) of data acquisition by reducing the 
computing demands for cross-correlations. Furthermore, this allows 
us to work with a relatively simple interferometric imaging 
algorithm without the concern of the so-called ``w'' term.  

In the early commissioning observations between 2005-2009, many efforts
were made to understand the performance of the 21CMA system including
antennas, receivers, data acquisition process, and even the power supply. 
We have developed all software packages and techniques such as identification
and mitigation of radio frequency interference (RFI), calibration, 
deconvolution, wide-field imaging, etc. necessary for full data
processing of 21CMA data. Since July of 2010 after its upgrade, 
the 21CMA has routinely observed the low frequency sky round the NCP, 
delivering about 2 terabits of visibility data per day for offline 
analysis. 

While the primary goal of the 21CMA is to accumulate a deep 
observation for EoR detection, even with relatively short integrations
we can characterize the extra-galactic source population. On one hand, 
these bright extragalactic sources, a byproduct of our EoR experiment, 
can be of great interest for exploration of their astrophysical 
properties. On the other hand, these radio sources constitute one 
of the major contamination components for our EoR detection, and 
therefore should be perfectly imaged and subsequently subtracted. 
In the current work we present the point radio sources observed 
with the 40 pods of the 21CMA E-W baselines for an integration of 
12 hours made on 2013 April 13. An extra deep sample with a 
higher sensitivity from a longer integration time of up to years 
will be published later.   

 \subsection{RFI Removal}

\begin{figure}
\begin{center}
\includegraphics[width=5cm, angle=-90]{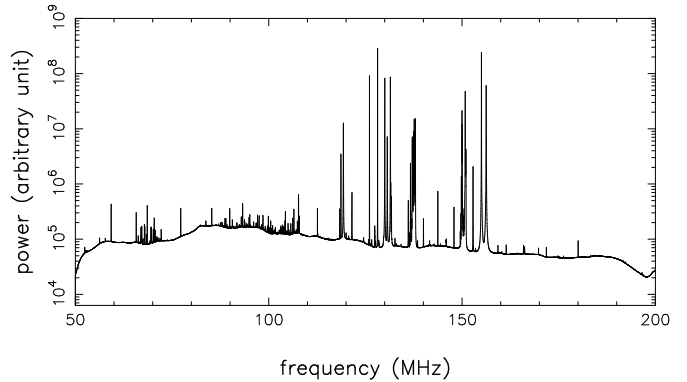}
\caption{A typical example of the average spectrum of visibility between 
two pods (E20 and W20, separated by a baseline of 2740 m) integrated 
over 12 hours. The frequency resolution is 24.4 kHz. }
\label{fig:spectrum}
\end{center}
\end{figure}

Although the radio environment of the 21CMA site, Ulastai, is exceptionally
quiet at low frequency and comparable to sites such as the
Murchison Radio Observatory (\citealt{Offringa15}), RFI is still a prominent 
concern for doing the EoR experiment. Figure \ref{fig:spectrum} displays the 
average spectrum of cross-correlation between two pods over 12 hours with 
a frequency resolution of $\Delta\nu=24.4$ kHz, which reveals the RFI at 
the site more clearly than the self-correlation of the same pod because 
thermal noises from different pods are not correlated. The overall 
shape reflects the decreasing sky noise at high frequencies and the 
decreasing efficiency of the antennas at low frequencies with a turnover 
at roughly 85 MHz. The strongest RFI sources are identified as low 
orbiting satellites at 137 MHz, local train communications at 150 MHz, 
civil aviation aircraft around 130 MHz and at 119 MHz, FM radio 
broadcasting at 88-108 MHz scattered by meteor tails and aircraft, 
and AM radio broadcasting around 70 MHz, and several other contaminators 
\citep{Huang16}. All these sources of RFI are highly time variable, 
and can be easily flagged for blanking during the editing process. 

After the mitigation of strong, time variable RFI, we apply a second 
statistical algorithm for further removal of other possible RFI using the 
fact that the sampling of visibilities should exhibit a Gaussian distribution 
for a given frequency, a scheme often adopted in data analysis of recent 
low frequency experiments (e.g. \citealt{Bowman07, Ghosh11}). In
the second stage of RFI mitigation we excluded 
all the data points beyond $3\sigma$ deviation from Gaussian statistics. 
This operation is applied to the real and imaginary parts separately, and the 
visibility data are edited out if either of the parts fail to reach
the $3\sigma$ restrictions. This still allows us to use the bulk (about 97\%) 
of the data except for some of the most noisy channels around 137 MHz 
and 150 MHz. The reader is referred to \citet{Huang16} for a detail 
description of our RFI identification and mitigation processing.

\subsection{Self-calibration}

Low-frequency radio interferometric observations are often affected 
by ionospheric turbulence, resulting in fluctuations of both source 
positions and flux densities. We run two algorithms to correct the phase error 
in the visibilities due primarily to ionospheric effects: self-calibration 
and the closure relation. Software developed in-house is used for 
self-calibration, in which a time interval of 1 hour is used for the present
study. We have tested shorter intervals of down to 10 minutes, and
found that the changes are only minor when longer baselines of 
$L>1500$ m are excluded (see below). We have already improved our 
self-calibration algorithm by combining conventional self-calibration
and redundant calibration, we will apply it for all 
baselines using a short time interval in the future data reduction (Zheng et al. in preparation). 
The NCP sky at 50-200 MHz is actually dominated by two bright sources: 
a central bright quasar (NVSS J011732+892848) and a galaxy 3C61.1. 3C61.1 is a very 
bright giant galaxy with a flux density of 33 Jy at 150 MHz, which also 
shows a complex structure. We select 10 radio sources including these 
two bright sources (Table \ref{table:Calibrators}) to correct the phase 
errors with self-calibration. Following this, we use the closure phase 
relation to check, and modify if necessary, the results.   
\begin{deluxetable*}{lllll}
 \tabletypesize{\scriptsize}
  \tablecaption{{10 sources used in the sky model for self-calibration\\
(Data are taken from NASA/IPAC EXTRAGALACTIC DATABASE (NED))\label{table:Calibrators}}}
  \tablehead{
 \colhead{Object Name}&
\colhead{$R_0$\tablenotemark{a}} &
\colhead{$S_{150{\rm MHz}}$\tablenotemark{b}}&
\colhead{$\alpha$\tablenotemark{c}}  &
\colhead{Source Type}\\
\colhead{}&
\colhead{(degree)}&
\colhead{(Jy)}&
\colhead{}&
\colhead{}
  }
  \startdata
NVSS J011732+892848 & 0.52 &  5.36$\pm$0.91 & 0.01$\pm$0.02 &point source\\
NVSS J062205+871948 & 2.67 &  4.38$\pm$0.49 &-0.86$\pm$0.02&point source\\ 
NVSS J133218+865005 & 3.17 &  2.94$\pm$0.33 &-0.98$\pm$0.02 &point source\\
NVSS J092016+862845 & 3.52 &  4.04$\pm$0.44 &-0.84$\pm$0.02 &point source\\ 
3C061.1              & 3.71 &  33.23$\pm$7.82 &-0.80$\pm$0.05 &resolved\\
NVSS J122518+860839  & 3.86 &  2.51$\pm$0.29 &-0.69$\pm$0.02 &point source\\
NVSS J1013.7+8553   & 4.12 &  3.01$\pm$0.34 &-0.77$\pm$0.02 & point source\\
NVSS J190350+853648 & 4.39 &  4.60$\pm$0.52 &-0.72$\pm$0.02 & point source\\
NVSS J194136+850138  & 4.97 &  4.53$\pm$0.71 &-0.86$\pm$0.03 &point source\\
NVSS J212926+845326  & 5.11 &  5.68$\pm$0.74 &-0.70$\pm$0.03 &point source
 \enddata
\tablenotetext{a}{The angular distance from the NCP.}
\tablenotetext{b}{Flux density measured at 150 MHz}
\tablenotetext{c}{The best-fit spectral index in terms of 
                    available data in NED.}

 \end{deluxetable*}

\subsection{Imaging}

To reduce further the ionospheric effect on our interferometric imaging
and improve image quality, we exclude both longer baselines of 
$L>1500$ m and the shorter ones of $L<100$ m in the present analysis. 
We perform a fast Fourier transform of the uv map using a uniform weighting
with $4096^2$ grids, sampled by different baselines and snapshots for each frequency channel, 
to generate the dirty maps. This yields a total of 6144 images 
over 50-200 MHz frequency bands with a frequency resolution of 24.4 kHz. 
Then, each dirty image is deconvolved using the conventional H\"{o}gbom 
CLEAN algorithm \citep{hog74} with a loop-gain of 0.05.
We terminate the CLEAN process when the fractional change in the total 
CLEANed flux density is less than $10^{-4}$ in the iteration.

Employment of a smaller threshold does not alter 
the result significantly. After the deconvolution we adopt a larger 
frequency bin of 1.56 MHz to combine the 6144 images into 96 mosaics, 
and concentrate on only the central field of view of $1024\times1024$ pixels, 
corresponding to a $14^2$ degree square area. Note that this gives rise 
to an angular resolution of $\sim 1$ arcmin per pixel while the best 
resolution of the array is only $3$ arcmin at 200 MHz.

\subsection{Calibrations of Primary Beam and Gains}
Directivity and gain of the 21CMA antenna element, the log-periodic antenna,
has been precisely measured in laboratory, which indeed allows us to 
compute the primary beam and gain of the 21CMA pod according to the 
pattern multiplication theorem. However,  in situ calibration has to be 
made to account for the possible effect of crosstalk between the 
antennas themselves and the antennas and the ground.
For a 12 hour (or more) observation of the 21CMA centered on the NCP, the 
primary beam of a pod has circular symmetry. Moreover, in terms of our 
laboratory measurement of the spatial response of the 21CMA log-periodic 
antenna and theoretical prediction based on the pattern multiplication 
theorem, the primary beam pattern of a 21CMA pod is nicely approximated 
by a Gaussian profile characterized by the standard deviation 
$\theta_b=3^{\circ}.62(\nu/100{\rm MHz})^{-1}$. Guided by our knowledge 
and experience, we adopt a modulated Gaussian function parametrized by 
\begin{equation} 
F(\theta,\nu)=G(\nu) (1+A\theta+B\theta^2) exp(-\theta^2/2\theta_b^2)
\end{equation}
for the primary beam and gain of the 21CMA pod, where G is the system 
gain which varies with frequency $\nu$, $\theta$ measures the 
radial distance from the NCP,  $\theta_b$ denotes the standard deviation 
of the Gaussian distribution, and A and B are the coefficients in the 
polynomial of degree 2. 

Unfortunately, there are no standard radio sources in the NCP region that 
can be used for calibration purposes. Thus we use the same sources 
selected in the self-calibration (Table \ref{table:Calibrators}) as our 
flux calibrators. We collect the flux density measurements of these sources made 
at different frequencies ranging from a few ten MHz to a few GHz from the
literature and obtain the spectral index, $\alpha$, for each source by 
fitting the observed flux densities to a power law, $S_\nu=S_{0}\nu^{\alpha}$. 
Except for the flat-spectrum sources, the fitting errors in amplitude
$S_{0}$ and spectral index  $\alpha$ are typically $\sim10\%$ and $\sim3\%$, 
respectively. Errors in these calibrators will be included in fitting of 
gain and spatial response function of the telescope. 
    
We then use the total flux density within a frequency bin of 1.56 MHz to 
accommodate our CLEANed images over the same frequency band. However,the 
primary beam is frequency-dependent, and varies roughly as an exponential 
profile characterized by $\theta_b$ for the 21CMA. This reduces further the 
number of the bright radio sources that can be used for calibration in the 
higher frequency bands. As a result, only the inner 4 sources within $3^{\circ}.52$ 
can be used for the  32 higher frequency bands with $\nu\geq 150$ MHz, for 
which uncertainties in the determination of gain and spatial response might 
be large. 

\begin{figure}
\begin{center}
\hspace{-4mm}
\includegraphics[width=5.5cm, angle=-90]{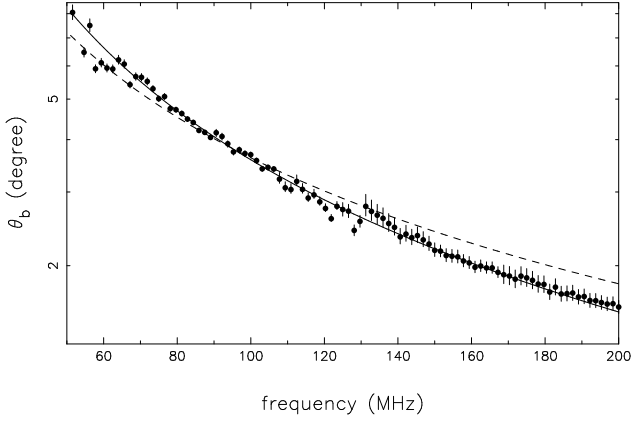}
\vspace{3mm}
\caption{The primary beam of the 21CMA pod, characterized by
the standard deviation of the Gaussian distribution $\theta_b$. 
The dots with error bars are the fitting results from all the 
sub-band images with a frequency width of 1.56 MHz. The solid curve 
presents the best-fit, $\theta_b=3^{\circ}.58(\nu/100{\rm MHz})^{-1.21}$,  
compared with the theoretically predicted variation of  
$\theta_b=3^{\circ}.62(\nu/100{\rm MHz})^{-1}$ (dashed).}
\vspace{3mm}
\label{fig:theta}
\end{center}
\end{figure}

\begin{figure}
\begin{center}
\hspace{-4mm}\includegraphics[width=8cm, angle=-90]{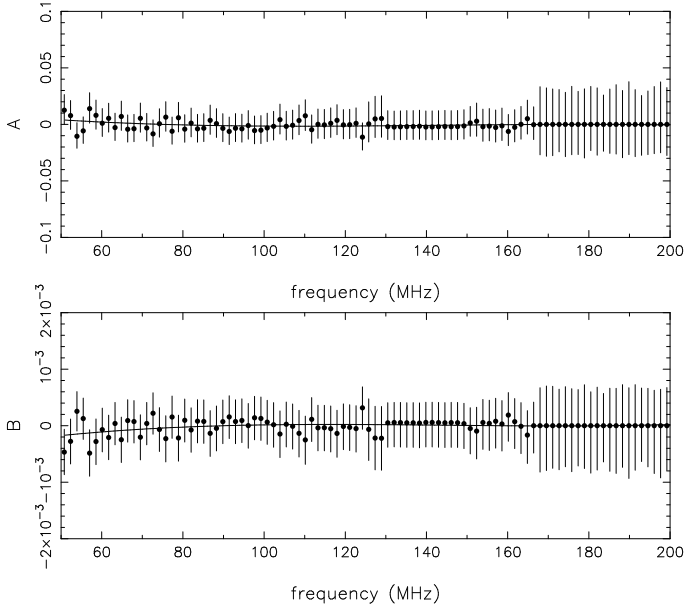}
\vspace{3mm}
\caption{The fitting results of the coefficients $A$ (top panel) and 
$B$ (bottom panel) in the polynomial of degree 2 to modify the Gaussian 
primary beam. The solid curve in each panel presents the best-fit, which 
is very close to zero.}
\vspace{3mm}
\label{fig:poly}
\end{center}
\end{figure}

\begin{figure}
\begin{center}
\hspace{-4mm}\includegraphics[width=6cm, angle=-90]{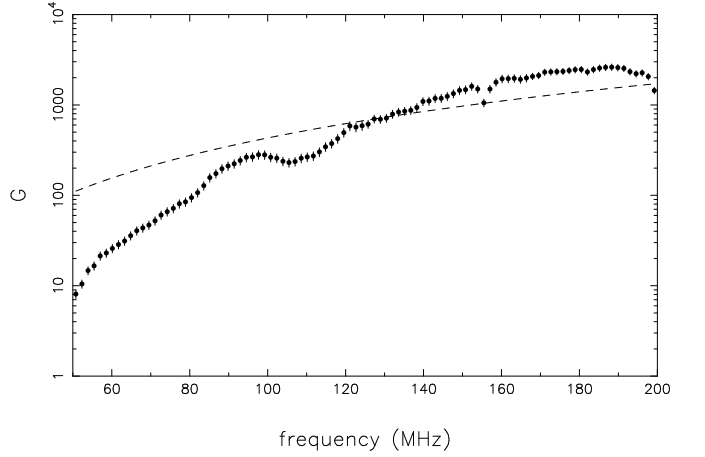}
\vspace{3mm}
\caption{Gain dependence of a 21CMA pod on frequency. Dots represent the 
best-fit result from the sub-band images with a frequency width of 1.56 MHz.
The dashed line shows the gain variation, $G\sim \nu^{2}$, for a single antenna 
element in terms of our design in order to guarantee a constant effective 
area over 50-200 MHz.}
\vspace{3mm}
\label{fig:G}
\end{center}
\end{figure}

We start with a Gaussian profile of 
$F(\theta,\nu)=G(\nu)exp(-\theta^2/2\theta_b^2)$, and find the best-fit
parameters $G$ and $\theta_b$ in each frequency channel. By fixing 
$G$ and $\theta_b$, we then use the same calibrators to find the 
modification parameters $A$ and $B$ to the Gaussian distribution.    
It turns out that the Gaussian profile provides rather a good 
description for the 12-hour averaged primary beam, and the 
modification parameters A and B are actually around zero 
(see Figures  \ref{fig:theta} and \ref{fig:poly}). Moreover, 
the best-fit Gaussian beam of
\begin{equation} 
\theta_b =3^{\circ}.58 \left(\frac{\nu}{100{\rm MHz}}\right)^{-1.21}
\end{equation}
is in good agreement with our theoretical prediction. Figure \ref{fig:G} 
shows the resultant system gain against frequency. The design of the 
21CMA antenna is such that the effective area remains roughly constant 
over a wide frequency range, which yields a frequency dependent gain of 
$G\propto \nu^{2}$. The steeper profile of the fitted gain factor also 
reflects the compensation for signal attenuation of the phased-delayed 
coaxial cable used for signal combination. 

\subsection{Dynamical Range}
\begin{figure}
\begin{center}
\hspace{-4mm}\includegraphics[width=9cm]{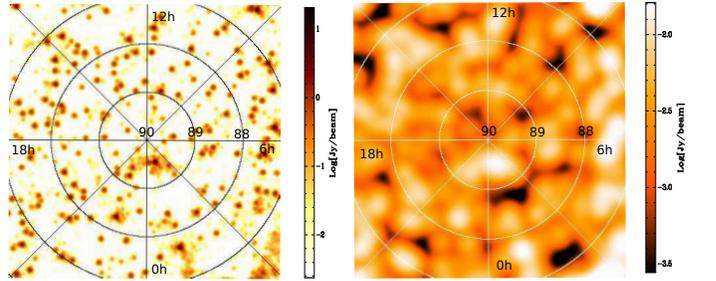}
\vspace{3mm}
\caption{An example of the CLEANed image (left panel) and 
corresponding noise map (right panel)
around the central region of the NCP at the subband 131.25 MHz, 
with an annulus of 1 degree in radial distance. }
\vspace{3mm}
\label{fig:noise}
\end{center}
\end{figure}

\begin{deluxetable}{lll}
  \tablecaption{{Image noise and beam size in the 12.5 MHz subband\\
\label{table:noise}}}
  \tablehead{
 \colhead{central frequency}&
\colhead{image noise} &
\colhead{beam size}\\
\colhead{(MHz)}&
\colhead{(mJy/beam)}&
\colhead{(arcmin)}
  }
  \startdata
81.25 & 28.9 &  6.78 \\
93.75 & 22.5 &  5.93\\
106.25 & 17.8 & 5.28 \\
118.75 & 10.7 & 4.75 \\
131.25 & 7.7 & 4.32 \\
143.75 & 5.3 & 3.96 \\
156.25 & 4.1 & 3.65 \\
168.75 & 3.3 & 3.39
 \enddata
 \end{deluxetable}

\begin{figure}
\begin{center}
\hspace{-4mm}\includegraphics[width=7cm, angle=-90]{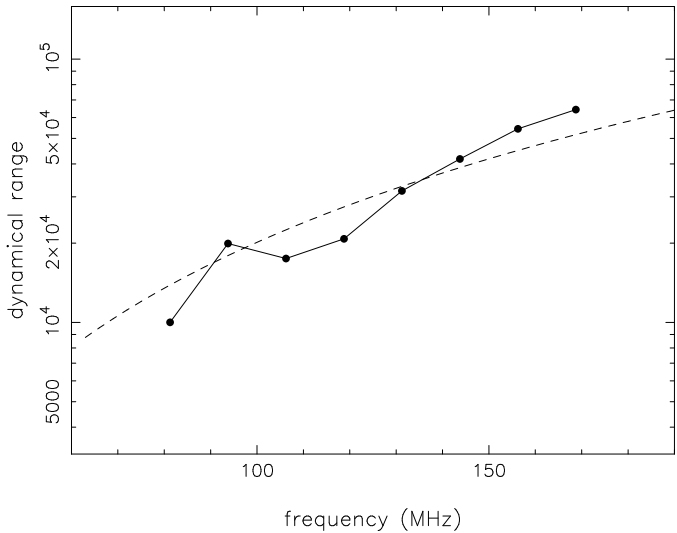}
\vspace{3mm}
\caption{Variation of dynamical ranges with frequencies. The dots denote 
the estimated ratio of $S_{\rm max}/S_{\rm noise}$ from the 8 sky maps centered at 
81.25, 93.75, 106.25, 118.75, 131.25, 143.75, 156.25 and 168.75 MHz, 
respectively. The dashed line represents the theoretically expected 
variation given by the flux density ratio of the cosmic radio sources 
($S_\nu\propto\nu^{-0.8}$ ) to the Milky Way as the noise 
($S_{\rm noise}\propto T_{\rm sky}\propto \nu^{-2.6}$), 
for which we have used an arbitrary amplitude. }
\vspace{3mm}
\label{fig:StoN}
\end{center}
\end{figure}

The CLEANed image in each of the 96 frequency channels is corrected 
for beam response and flux gain. Then the images are stacked into 12 
sky maps with a bandwidth of 12.5 MHz. To reduce further the possible 
roll-off effect of the bandpass filters used in the 21CMA receivers, 
we will not use the data at the two frequency ends, 50-75 MHz and 
175-200 MHz, in the present paper. This yields a total of 8 sky maps 
centered at 81.25, 93.75, 106.25, 118.75, 131.25, 143.75, 156.25 and 
168.75 MHz, respectively, with a varying image noise from $\sim 28.9$ mJy 
at lower frequency end to  $3.3$ mJy higher frequency one. 
Figure \ref{fig:noise} illustrates an example of the CLEANed image 
and corresponding noise map in the 131.25 MHz subband.
The noise and beam size in each 12.5 MHz subband image
are summarized in Table \ref{table:noise}. We calculate 
the dynamical range of each sky map to provide a quantitative estimate 
of the image quality, which we defined here as the ratio of the peak brightness 
$S_{\rm max}$ of the image to the rms noise $S_{\rm noise}$, and the result 
is shown in Figure \ref{fig:StoN} for the 8 sky maps. It appears that 
the ratio of $S_{\rm max}/S_{\rm noise}$ reaches a value of $10^4-6\times10^4$ 
and shows an increase trend with frequency. Such a variation is simply 
the combined effect of the system sensitivity, which is dominated by the 
Milky Way (e.g. \citealt{Holder12}) 
$S_{\rm noise}\propto T_{\rm sky}\propto \nu^{-2.6}$, 
and the power-law behavior of the background radio source flux density, 
$S_\nu\propto\nu^{\alpha}$ with $\alpha\approx-0.8$. As a consequence, 
the dynamical range of the sky map varies with frequency roughly as 
$S_{\rm max}/S_{\rm noise}\propto \nu^{1.8}$. Yet, the current dynamical 
range is still one order of magnitude below the EoR detection requirement
(see \citealt{Barkana01}; \citealt{Furlanetto06}; \citealt{Pritchard10} for
review), and we will address the possible reasons in the sub-section on noise 
analysis (Section 3.5) below. We have also made a full band 75-175 MHz
image by mosaicing the 8 sub-band images together after convolving to
a common resolution of $\sim 6$ arcmin (Figure \ref{fig:dirtymap} 
and Figure \ref{fig:wideband}). This combined image will be used to 
remove spurious sources (see below).

\section{Source Catalog and Properties}

\subsection{Source Selection}

The field of view (FoV) depends on the observing frequency, such that 
FoV$\propto\nu^{-2}$. As a result, we can only extract full spectral 
information for radio sources within a radius of $\sim3^{\circ}$ around 
the NCP. Nonetheless, we also count the sources between $3^{\circ}-5^{\circ}$ 
to expand our radio source catalog, for which spectral measurements
are based on only the four lower frequency channels, 81.25, 93.75, 106.25 
and 118.75 MHz. We note that, many more sources have been found beyond 
$5^{\circ}$ at the lower frequency end, but we will not include them in
the present catalog. 

\begin{figure}
\begin{center}
\hspace{-4mm}\includegraphics[width=9cm, angle=0]{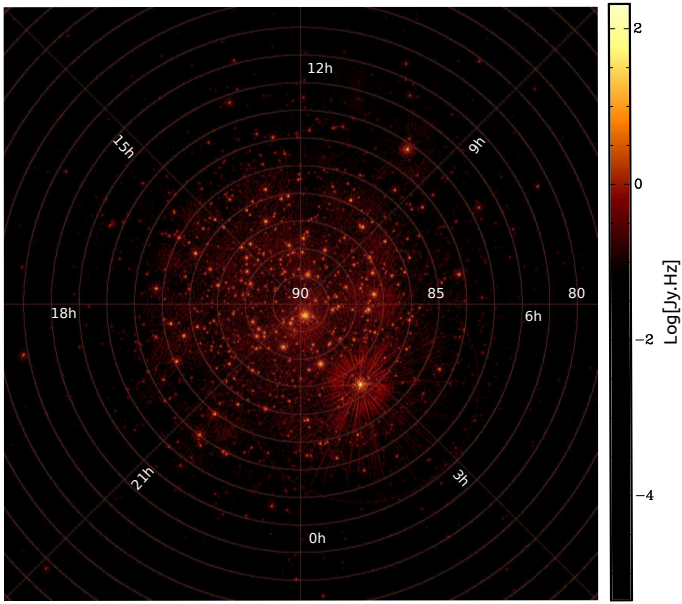}
\vspace{3mm}
\caption{The dirty map centered on the NCP in the frequency range of 
75-175 MHz. A larger field around the NCP is shown, with an annulus 
of 1 degree in radial distance. }
\vspace{3mm}
\label{fig:dirtymap}
\end{center}
\end{figure}

\begin{figure}
\begin{center}
\hspace{-4mm}\includegraphics[width=9cm, angle=0]{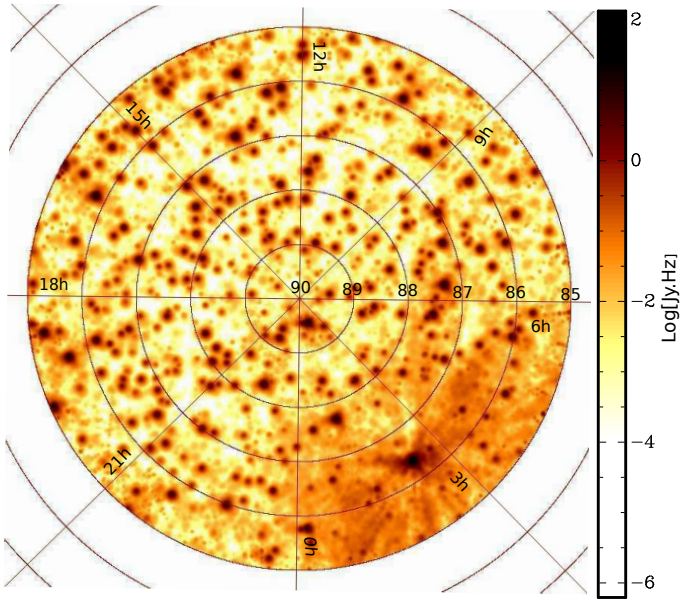}
\vspace{3mm}
\caption{The restored sky map in the frequency range of 75-175 MHz.
A central area of $5^{\circ}.5\times5^{\circ}.5$ is shown, with an annulus
of 1 degree in radial distance. The deconvolution is performed through
the H\"ogbom CLEAN algorithm with a loop gain of 0.05. The brightest 
source (lower right), 3C061.1, is located at a distance of $3^{\circ}46'$ 
from the NCP, and deconvolution noise around 3C061.1, which arises mainly 
from grating lobes and inaccurate calibration, is clearly visible. Note 
that we have enhanced the image contract to display the noise field 
which is actually four orders of magnitude fainter than the typical 
bright sources in the field.}
\vspace{3mm}
\label{fig:wideband}
\end{center}
\end{figure}

Source candidates are first selected in the full-band image covering 
75-175 MHz, using a method similar to DUCHAMP \citep{Whiting01}, in
which a threshold of $3\sigma$ is adopted to identify source candidates. 
The position of the peak flux of each candidate in the full-band image 
is tagged and then used for selection of source candidate in other 
sub-bands in a procedure similar to that described in \citet{NHW14}. 
The source candidates will be removed from the list if 
their counterparts are missing in one of the frequency bands. This 
allows us to identify and thus remove all the spurious sources that 
result from the sidelobes of the brightest radio source, 3C061.1 
(see Figure \ref{fig:dirtymap} and Figure \ref{fig:wideband}). Indeed, 
we are still unable to perfectly CLEAN the sidelobes of the brightest 
radio galaxy, 3C061.1, located at $3^{\circ}.71$ degrees from NCP. The 
complex structure of this source \citep{Lawrence96} is marginally 
resolved with 21CMA. Even with a better angular resolution and a 
multi-directional calibration using SAGECAL \citep{Yatawatta09,Kazemi11}, 
LOFAR imaging of the same source also contains significant errors 
\citep{Yatawatta13}. Our current experiment towards the detection
of the EoR is thus largely limited by the presence of this very bright, 
complex radio galaxy in the NCP field.

A Gaussian profile is used to fit the ``surface brightness'' of a radio 
source in each of the sub-bands and the corresponding flux density can be computed 
straightforwardly. When two sources show an overlap due to poor resolutions, 
a two-component Gaussian profile is adopted to fit the images simultaneously. 
Yet, even this double Gaussian models may fail to fit two sources with 
significant overlap especially at low frequencies. In this case, we will 
not provide the flux density information for the two sources but still treat them 
as separate sources. We calculate the spectral index for each source using the 
measured flux densities in all the available bands, up to 8 bands for sources 
within $3^{\circ}$ and up to 4 bands for the ones in the annulus of 
$3^{\circ}-5^{\circ}$.  

\begin{figure}
\begin{center}
\hspace{-4mm}\includegraphics[width=9cm, angle=0]{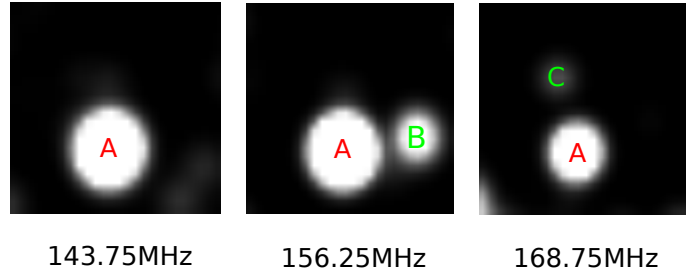}
\vspace{3mm}
\caption{An example of source selection based on adjacent channels:
'A' is selected as a cosmic source in the subband 156.25 MHz
because it occurs at the same position in at least one of the adjacent 
channels (143.75 MHz and 168.75 MHz). On the contrary, neither 'B'
nor 'C' will be classified as source candidate because they show up
in a single frequency channel only.}
\vspace{3mm}
\label{fig:sourceselection}
\end{center}
\end{figure}

Our method of source identification is indeed reliable for mitigation of 
artificial sources. However, the fainter sources below our threshold at 
least in one of the eight sub-bands can not be included with the current 
criterion of source selection. In addition, the threshold varies with frequency 
channels, and therefore, we may have missed a significant fraction of faint 
sources. We have thus tested two other methods by relaxing the above 
criterion for source selection. The first is to work with an extreme 
case: all the candidates with peak fluxes above $3\sigma$ in any sub-band 
are selected. This provides an overestimate of source population because image 
artifacts, due to imperfect CLEANing of sidelobes of bright sources, especially 
3C061.1, have contaminated our observing field. We demonstrate in Section 3.3 
an example of the completeness estimate from such a source selection 
method for the highest angular resolution sub-band at 168.75 MHz. 
It turns out that the contamination of spurious sources is 
indeed very serious. The second test is to select sources based on 
adjacent frequency channels instead of full bands. To deal with the sidelobes 
from the bright sources, source candidates should be removed from 
the list once their counterparts are missing only in one of the adjacent 
channels, using the frequency-dependence properties of sidelobes 
(Figure \ref{fig:sourceselection}). 
This allows us to identify more faint sources yet leaves the following two 
drawbacks: (1) the position variations of the sidelobes-induced ``sources'' 
become indistinguishable at low frequencies because of poor angular 
resolution, and (2) the spectral index cannot be estimated reliably if 
a source shows up only in two frequency bands. While we adopt our 
conservative way to select sources and construct our source catalog 
in this work, we will compare the completeness and source counts among 
the three selection methods below and will improve our source selection 
algorithm in our future work.

\begin{deluxetable*}{ccccccccccc}
 \tabletypesize{\scriptsize}
 \tablecaption{Source Catalog. 
The coordinates, total fluxes at eight frequency bands (if available) 
and spectral indices are provided for 624 sources detected within $3^{\circ}$
(8 sub-bands) and in annulus of $3^{\circ}-5^{\circ}$ (4 lower sub bands)
around the NCP. The positional accuracy of the central coordinates is
 $\sim5$ arcmin. Table 3 is published in its entirety in the electronic 
edition of MRT-sourcecatalog.txt, and a portion is shown here for 
guidance regarding its form and content. 
\label{table:list}}
\tablehead{
\colhead{RA}&
\colhead{DEC}&
\colhead{$F_{\rm 81.25}$}&
\colhead{$F_{\rm 93.75}$}&
\colhead{$F_{\rm 106.25}$}&
\colhead{$F_{\rm 118.75}$}&
\colhead{$F_{\rm 131.25}$}&
\colhead{$F_{\rm 143.75}$}&
\colhead{$F_{\rm 156.25}$}&
\colhead{$F_{\rm 168.75}$}&
\colhead{$\alpha$}\\
\colhead{(J2000)}&
\colhead{(J2000)}&
\colhead{(Jy)}&
\colhead{(Jy)}&
\colhead{(Jy)}&
\colhead{(Jy)}&
\colhead{(Jy)}&
\colhead{(Jy)}&
\colhead{(Jy)}&
\colhead{(Jy)}&
\colhead{}\\
\colhead{(deg)}&
\colhead{(deg)}&
\colhead{}&
\colhead{}&
\colhead{}&
\colhead{}&
\colhead{}&
\colhead{}&
\colhead{}&
\colhead{}&
\colhead{}
}
\startdata
  9.49 &   89.68 &  $0.033^{+0.005}_{-0.005}$ &   $0.031^{+0.003}_{-0.003}$ & $0.026^{+0.003}_{-0.003}$   &  $0.025^{+0.003}_{-0.003}$ & $0.013^{+0.002}_{-0.002}$ &  $0.007^{+0.001}_{-0.001}$ &  $0.015^{+0.003}_{-0.003}$ & $0.021^{+0.004}_{-0.004}$ & $-1.37^{+0.20}_{-0.20}$ \\
  5.23 &   89.65 &  $0.391^{+0.055}_{-0.055}$ &  $0.287^{+0.029}_{-0.029}$ & $0.303^{+0.032}_{-0.032}$ &  $0.339^{+0.037}_{-0.037}$ &  $0.302^{+0.047}_{-0.047}$ &  $0.218^{+0.025}_{-0.025}$ &  $0.195^{+0.036}_{-0.036}$ &  $0.184^{+0.038}_{-0.038}$ &  $-0.93^{+0.19}_{-0.19}$ \\
  7.59 &   89.63 &   \nodata   &   $0.044^{+0.005}_{-0.005}$ &  $0.033^{+0.004}_{-0.004}$   &  $0.026^{+0.003}_{-0.003}$ &   $0.030^{+0.005}_{-0.005}$ &   $0.027^{+0.003}_{-0.003}$ &   $0.038^{+0.007}_{-0.007}$ &   $0.053^{+0.011}_{-0.011}$ &   $0.22^{+0.19}_{-0.19}$ \\
  0.29 &   89.58 &   \nodata   &   $0.222^{+0.023}_{-0.023}$ &   $0.245^{+0.026}_{-0.026}$ &  $0.230^{+0.025}_{-0.025}$   &   $0.226^{+0.035}_{-0.035}$ &   $0.271^{+0.031}_{-0.031}$ &   $0.247^{+0.046}_{-0.046}$ &   $0.185^{+0.038}_{-0.038}$ &  $-0.11^{+0.20}_{-0.20}$ \\
  12.30 &   89.57 &   \nodata   &  \nodata   &  \nodata   &   $0.054^{+0.006}_{-0.006}$ &   $0.052^{+0.008}_{-0.008}$ &   $0.047^{+0.005}_{-0.005}$ &   $0.049^{+0.009}_{-0.009}$ &   $0.054^{+0.011}_{-0.011}$ &  $-0.09^{+0.49}_{-0.49}$ \\
  16.55 &   89.55 &   $0.512^{+0.072}_{-0.072}$ &   $0.443^{+0.045}_{-0.045}$ &   $0.548^{+0.058}_{-0.058}$ &   $0.584^{+0.063}_{-0.063}$ &   $0.451^{+0.070}_{-0.070}$ &   $0.417^{+0.047}_{-0.047}$ &   $0.322^{+0.060}_{-0.060}$ &   $0.273^{+0.056}_{-0.056}$ &  $-0.75^{+0.16}_{-0.16}$ \\
  1.29 &   89.47 &   \nodata   &  \nodata   &   \nodata   &   \nodata   &   $7.272^{+1.125}_{-1.125}$ &   $6.613^{+0.750}_{-0.750}$ &   $6.450^{+1.230}_{-1.230}$ &   $7.498^{+1.542}_{-1.542}$ & $0.09^{+0.56}_{-0.56}$\\
  9.52 &   89.47 &   \nodata   &  \nodata   &   $0.164^{+0.017}_{-0.017}$ &   $0.127^{+0.014}_{-0.014}$ &   $0.150^{+0.023}_{-0.023}$ &   $0.143^{+0.016}_{-0.016}$ &   $0.113^{+0.021}_{-0.021}$ &   $0.113^{+0.023}_{-0.023}$ &  $-0.69^{+0.29}_{-0.29}$ \\
  6.31 &   89.45 &   $1.101^{+0.155}_{-0.155}$ &   $0.896^{+0.092}_{-0.092}$ &   $0.800^{+0.085}_{-0.085}$ &   $0.585^{+0.063}_{-0.063}$ &   $0.586^{+0.091}_{-0.091}$ &   $0.466^{+0.053}_{-0.053}$ &   $0.431^{+0.080}_{-0.080}$ &   $0.374^{+0.077}_{-0.077}$ &  $-1.48^{+0.15}_{-0.15}$ \\
  10.86 &  89.40&    $0.072^{+0.014}_{-0.010}$ &   $0.066^{+0.007}_{-0.007}$ &   $0.071^{+0.008}_{-0.008}$ &   $0.048^{+0.005}_{-0.005}$ &   $0.070^{+0.011}_{-0.011}$ &   $0.059^{+0.007}_{-0.007}$ &   $0.073^{+0.014}_{-0.014}$ &   $0.057^{+0.012}_{-0.012}$ &  $-0.16^{+0.16}_{-0.16}$ 
\enddata
\end{deluxetable*}

\begin{figure}
\begin{center}
\hspace{-4mm}\includegraphics[width=7cm,angle=-90]{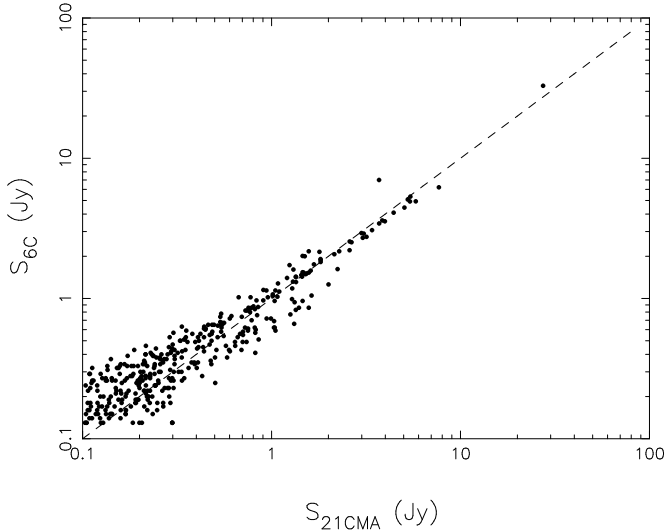}
\vspace{3mm}
\caption{A comparison of flux densities measured at 151 MHz in the 6C 
survey and the 21CMA observation above 0.1 Jy. Errors on the points
are omitted for clarity. The dashed line indicates equal flux 
density values.}
\vspace{3mm}
\label{fig:21cma-6c}
\end{center}
\end{figure}

A total of 624 radio sources are found around the NCP region, and 
their positions, integrated fluxes and spectral indices are listed in 
Table \ref{table:list}. 
Note that for sources within $3^{\circ}$ we fit integrated fluxes in 
eight subbands to a power-law to estimate their spectral indices.
While for sources in annulus of $3^{\circ}-5^{\circ}$ we only use 
the integrated fluxes measured in 4 lower frequency subbands.  
Moreover, data points in some of the subbands have not been taken if
the integrated fluxes of images cannot be precisely measured due to poor 
resolution and/or significant overlap. 
Of the 624 sources, 322 sources lie between 3 and 5 deg from the field
center NCP which are selected using the four lower frequency subbands
(81.25, 93.75, 106.25 and 118.75 MHz). 
An immediate cross-check with the existing 
the 6C catalog at 151 MHz reveals 490 counterparts. The missing sources 
are the fainter ones below the detection limit of 0.1 Jy in the 6C survey. 
Namely, about 134 new sources are detected around and below 0.1 Jy 
at 151 MHz. Figure \ref{fig:21cma-6c} compares the flux densities of these 
490 sources at 151 MHz listed in both 6C catalog and our new catalog, 
and two measurements above 0.1 Jy show rather a good agreement. 

\subsection{Spectral index}

\begin{figure}
\begin{center}
\hspace{-4mm}\includegraphics[width=7cm,angle=-90]{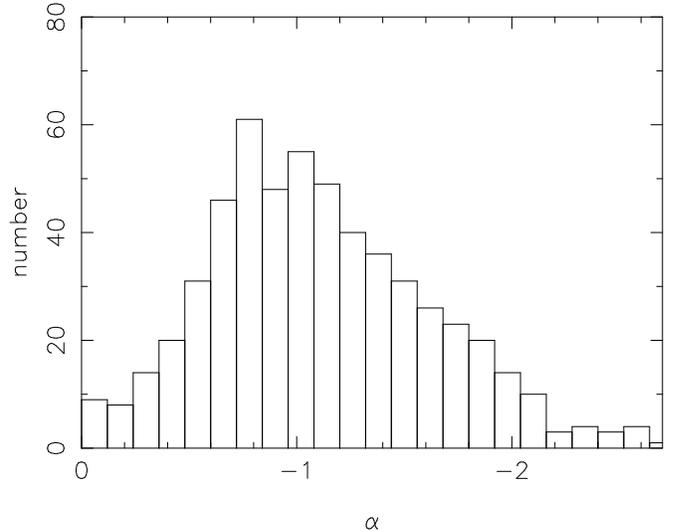}
\vspace{3mm}
\caption{Histogram of spectral indices $\alpha$ ($S\sim\nu^{\alpha}$) 
for the 21CMA source catalog.}
\vspace{3mm}
\label{fig:alpha1}
\end{center}
\end{figure}

\begin{figure}
\begin{center}
\hspace{-4mm}\includegraphics[width=6cm,angle=-90]{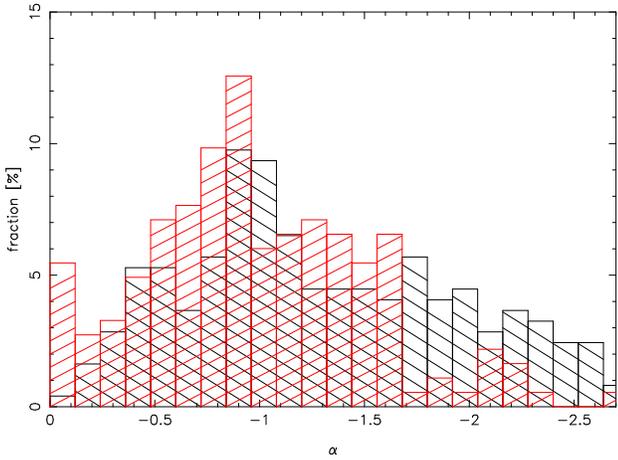}
\vspace{8mm}
\caption{Comparison of spectral indices derived between the high-frequency
subbands (black) and lower-frequency ones (red) within 3 degrees.
Note that the vertical axis denotes the source fraction}
\vspace{3mm}
\label{fig:alpha2}
\end{center}
\end{figure}

\begin{figure}
\begin{center}
\hspace{-4mm}\includegraphics[width=6cm,angle=-90]{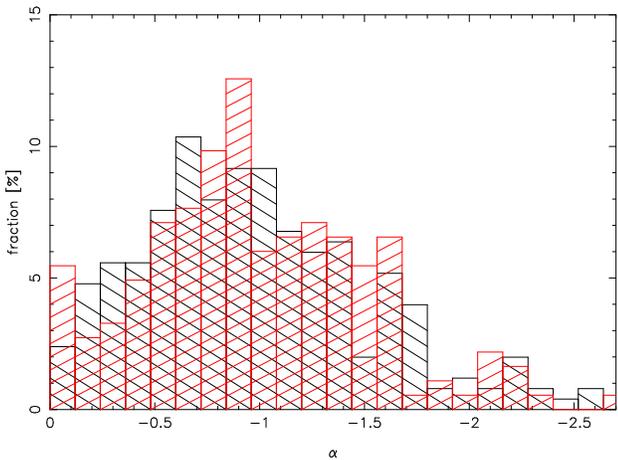}
\vspace{3mm}
\caption{Comparison of spectral indices for sources in the central region 
of  $3^{\circ}$ (red) around the NCP and in the outer annulus of 
$3^{\circ}-5^{\circ}$ (black). The vertical axis denotes 
the source fraction. }
\vspace{3mm}
\label{fig:alpha3}
\end{center}
\end{figure}

Figure \ref{fig:alpha1} illustrates the histogram of the spectral indices 
of the 624 sources presented here. While the distribution peaks at 
approximately $\alpha\approx-0.8$, consistent with what is expected for 
the overall population of cosmic radio sources, the asymmetrical behavior 
indicates that more sources tend to have a steeper spectral index. This 
could arise as the radio sources with steeper indices are more easily 
detected at low frequencies. In other words, low-frequency observations 
would probably miss some of the flat-spectrum sources if they are not 
bright enough. Yet, recent observations have argued a spectral flattening 
towards both lower frequencies and fluxes (e.g. \citealt{Intema11}; 
\citealt{Heywood13}; \citealt{vanWeeren14}). To investigate if similar behavior
is evident in our radio source catalog, we restrict the sources within 3 degrees and 
compare in Figure \ref{fig:alpha2} the spectral indices between the 4 lower 
frequency subbands and the 4 higher frequency subbands.  It turns out that 
there is a significant excess of steep-spectrum sources in the higher frequency 
subbands. This can be further quantified by considering the means, medians and their ratios 
(means/medians) of the two distributions, yielding (-1.25,-1.15,1.09) and
(-0.86,-0.87,0.99) for the higher and lower frequency subbands, respectively.  
We have also compared the low-frequency sources in the inner 
radius of 3 degrees and the ones in the outer annulus of $3^{\circ}-5^{\circ}$, 
and found that their spectral indices are essentially similar 
(Figure \ref{fig:alpha3}). While the absence of bias in spectral index 
between the sources within the central region ($\le3^{\circ}$) and the ones in 
the outer region ($3^{\circ}-5^{\circ}$) simply reflects  
the uniformity of cosmic radio sources,  it also indicates that 
our beam model is correct. It turns out that our analysis does support 
a flattening trend toward lower frequencies. If confirmed, this may 
add a further difficulty for low-frequency observations to detect faint, 
flat-spectrum sources.    

\subsection{Completeness}

\begin{figure}
\begin{center}
\hspace{-4mm}\includegraphics[width=7.5cm,angle=-90]{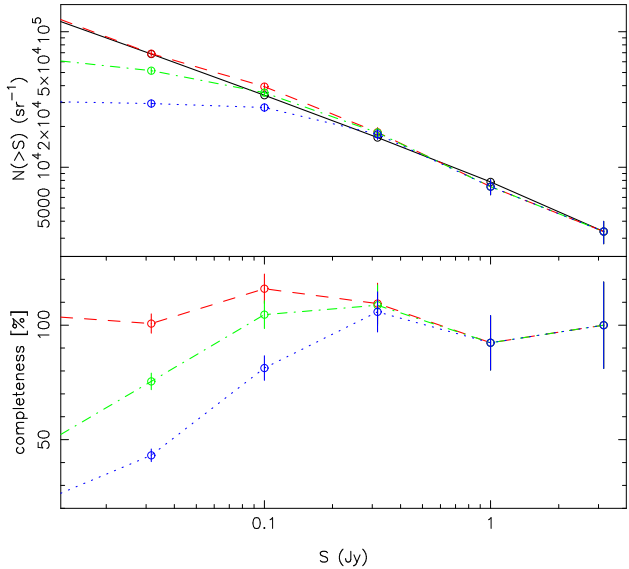}
\vspace{3mm}
\caption{The integrated source counts $N(>S)$ (top) and detection 
fractions (bottom)  estimated using the \citet{Intema11} deep survey 
as control field, $dN/dS=5370S^{-1.59}$ (The integrated form is plotted 
as a black line). Flux densities are converted to 168.75 MHz 
by assuming a power law with spectral index of $-0.8$, and only a 
central field with 3 degrees is chosen. (1) Red points: peaks 
above $3\sigma$ in any subbands are taken to be ``sources''; 
(2) Green points: source selection is based on their positions at two 
adjacent frequency bands; (3) Blue points: source should be visible 
at all eight sub-bands. }
\vspace{3mm}
\label{fraction-3methods}
\end{center}
\end{figure}

Before we perform  Monte-Carlo simulations to estimate the completeness 
of our source selections, we compare our source counts with the deep 
low-frequency observations of the NOAO Bo\"{o}tes field down to 6 mJy made 
with GMRT at 153 MHz \citep{Intema11}. This allows us to make a quick yet 
rough judgment on the incompleteness of our source counts at different 
fluxes. Of course, an extrapolation based on a spectral index conversion 
should be made for source counts at other frequencies for the ``control'' 
field. We notice that the reliability of such an extrapolation is probably 
questionable, as argued recently by \citet{Franzen16}. Taking the differential 
source count by \citet{Intema11} at 153 MHz, $dN/dS=5370 S^{-1.59}$, 
we calculate the ratios of our observed sources $N(>S)$ to what are expected 
in terms of this single power law, and plot the results for the 168.75 MHz 
frequency channel in Figure \ref{fraction-3methods}.
It appears that the incompleteness 
for our standard method of source selection occurs around 0.1 Jy, where the 
ratio drops to $\sim80\%$. Namely, a significant fraction of the fainter sources 
may be missed below 0.1 Jy when sources are required to be visible in all subbands 
to guarantee that the artificial sources from sidelobes be clearly discriminated 
and perfectly removed. We have also shown the completeness estimates for the
other two selection methods: The extreme approach that treats all the 
peaks above $3\sigma$ as ``sources'' gives rise to an apparent over-estimate 
of true sources even down to $\sim0.01$ Jy, indicating that the contamination 
is rather significant. Sources that are selected in terms of their locations
at adjacent frequency channels seem to be complete until $\sim0.1$ Jy,
and their completeness drops to $\sim50\%$ at 0.01 Jy. 
While the latter provides a ``better'' completeness of source selection, 
the poor angular resolution at lower frequencies may be inefficient for
discrimination of position shifts of the artificial sources due to sidelobes, 
as was pointed out above.     

\begin{figure}
\begin{center}
\hspace{-4mm}\includegraphics[width=9cm]{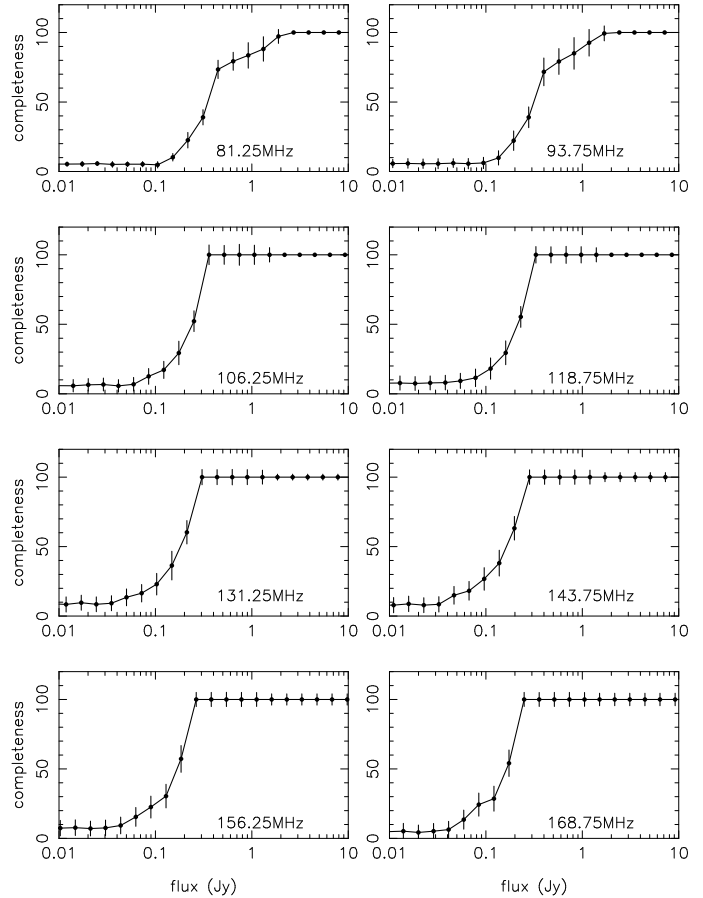}
\vspace{3mm}
\caption{The estimated completeness against total flux density 
for the eight frequency subbands. A total of 1600 artificial sources following
 $dN/dS\propto S^{-1.59}$ are randomly distributed over a circular field 
with a radius of 3 degrees, and noise and beam smearing are added in terms 
of the true measurements on the NCP field. The same source selection 
method as we adopted for the NCP field is applied to the simulation.
Error bars at each frequency represent the standard deviation 
from 20 realizations. }
\vspace{3mm}
\label{fraction}
\end{center}
\end{figure}

We now perform a Monte-Carlo simulation to quantify the completeness of
the source sample. We generate 20 fields with a radius of $3^{\circ}$, each 
of which contains 1600 point sources with positions randomly distributed. 
The fluxes of these artificial sources are assumed to follow a power-law 
of $dN/dS\propto S^{-1.59}$ and are further restricted within 
$5$ mJy $\leq S\leq30$ Jy, for which a spectral index of $-0.8$ is assumed.
We include the beam smearing effect by convolving the image with a Gaussian 
function characterized by the true beam at each frequency band. We add noise 
to the simulated map at each channel using the true residual map from our 
deconvolution after point sources are peeled out. Furthermore, the radial 
variation in sensitivity around the NCP is taken into account by the spatial 
response of the telescope described by Eq.(2). Now, we adopt the same 
mechanism, the so-called standard method, to extract sources at the same eight 
frequency channels, using the image flux and position information over 
the frequency bands. Figure \ref{fraction} shows the mean fraction 
of simulated sources recovered against total flux density for each of the 
eight channels, respectively, in which the error bars represent the standard 
statistical deviations. As expected, the turning points of completeness exhibit a 
weak increasing tread towards higher frequencies. Unfortunately, the completeness
decreases rapidly to zero beyond $S\sim0.1$ Jy, although we can still 
detect some sources down to $S\sim0.01$ Jy for all the 8 frequency bands. 
Our estimated  completeness will be used for correcting the 
incompleteness of source counts below.

\subsection{Source Counts}

\begin{figure}
\begin{center}
\hspace{-4mm}\includegraphics[width=8cm]{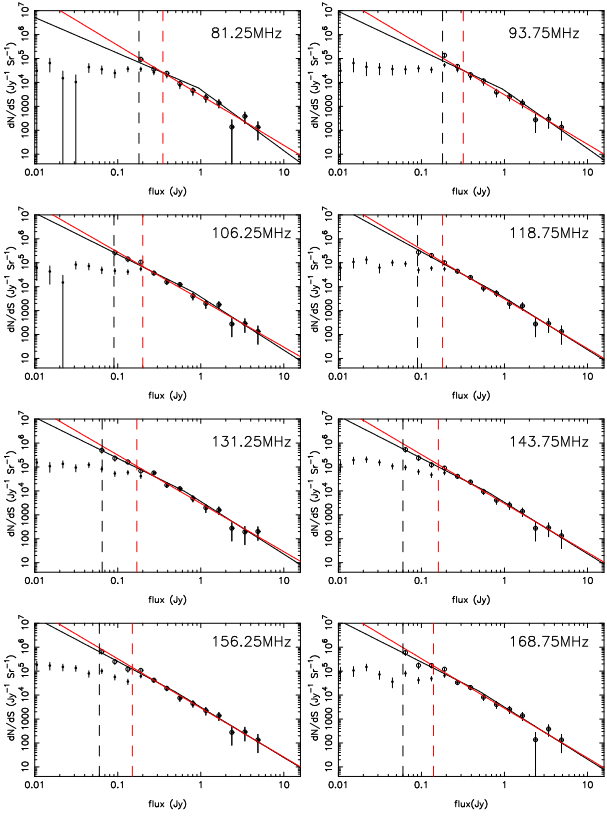}
\vspace{3mm}
\caption{Differential source counts over the central field within 
a radius $3^{\circ}$ for each of the eight frequency subbands. 
Filled and open circles represent the raw data and 
completeness-corrected data, respectively.
The two vertical dashed lines indicate the $20\%$ (black) completeness 
and $50\%$ (red) completeness, respectively.  
The best-fit two power laws for the $20\%$ completeness limit
are shown by the black solid lines, while the results for the $50\%$ 
completeness data are illustrated by the red solid lines. 
}
\vspace{3mm}
\label{numbercount}
\end{center}
\end{figure}

We plot in Figure \ref{numbercount} the differential source counts $dN/dS$ of 
our sample over the central field of radius $3^{\circ}$ for each of the 8 
frequency bands, in which both the raw data and completeness-corrected data 
in terms of the above Monte-Carlo simulation are displayed.
For data points with completeness of $50\%$ or above, which corresponds to 
the brighter sources with fluxes greater than $\sim0.1$ Jy,   
a single power law fits nicely the completeness-corrected data:
\begin{equation} 
\frac{dN}{dS}=k S^{-\gamma}{\rm Jy}^{-1}{\rm sr}^{-1}.
\end{equation}
If we adopt a lower completeness of $20\%$ to include more fainter
sources, though the data points below $50\%$  already become unreliable, 
$dN/dS$ would follow the the well-known two power laws for cosmic radio 
population at low frequencies \citep{Hales88,Cohen07,Moore13}.
\begin{equation} 
\frac{dN}{dS}=\left\{ 
  \begin{array}{ll}
    k_1 S^{-\gamma_1}{\rm Jy}^{-1}{\rm sr}^{-1}, & S<S_0;\\
    k_2 S^{-\gamma_2}{\rm Jy}^{-1}{\rm sr}^{-1}, & S\ge S_0
  \end{array} \right.
\end{equation}
where $S_0$ indicates the turnover flux. The best-fit parameters 
for the 8 frequency subbands are summarized in Tables \ref{table:Ncounta} and
\ref{table:Ncountb} for the $50\%$ and $20\%$ completeness data, respectively.
It appears that for the bright population with $S>S_0$, our best-fit indices 
vary between 2.10 and 2.55, which are both consistent with each other within error 
bars, and are also close to 2.5 which is expected for a Euclidean cosmology. 
For the fainter radio sources below $S_0$, we find a weak dependence of $\gamma_1$ 
on frequency, with a flattening source count towards lower frequencies. 
We note that a power-law fit at  $S<0.4$ Jy to the GMRT counts 
from Williams, Intema \& Rottgering (2013), Intema et al. (2011) and 
Ghosh et al. (2012) gives $\gamma_1=1.54$ at 154 MHz \citep{Franzen16} 
and the GMRT targeted survey at 153 MHz finds essentially similar 
results, $\gamma_1=1.59$ (\citealt{Intema11}, see also \citealt{Ghosh12}). 
Our best-fit power indices at the two closest bands, 143.75 and 156.25 MHz, 
are $1.77\pm0.06$ and $1.83\pm0.10$, respectively, which are slightly higher
than these results due to the lack of fainter sources in our samples.   

\begin{deluxetable}{lll}
  \tablecaption{The best-fitted parameters for differential source counts 
($50\%$ completeness). 
            \label{table:Ncounta}}
  \tablehead{
  \colhead{$\nu_c$}&
  \colhead{$k$}&
  \colhead{$\gamma$}\\
  \colhead{(MHz)}&
    \colhead{}&
  \colhead{}
  }
  \startdata
81.25 & 2896$\pm$898 & 2.09$\pm$0.22 \\
93.75 & 3044$\pm$737 & 2.06$\pm$0.21 \\
106.25& 2879$\pm$635 & 2.01$\pm$0.24 \\
118.75& 3062$\pm$979 & 2.07$\pm$0.32 \\
131.25& 3007$\pm$679 & 2.02$\pm$0.28 \\
143.75& 2974$\pm$789 & 2.05$\pm$0.30 \\
156.25& 2911$\pm$757 & 2.05$\pm$0.33 \\
168.75& 2812$\pm$464 & 2.07$\pm$0.28 
  \enddata
 \end{deluxetable}

\begin{deluxetable}{llllll}
  \tablecaption{The best-fitted parameters for differential source counts
($20\%$ completeness).
            \label{table:Ncountb}}
  \tablehead{
  \colhead{$\nu_c$}&
  \colhead{$S_0$}&
  \colhead{$k_1$}&
  \colhead{$\gamma_1$}&
  \colhead{$k_2$}&
  \colhead{$\gamma_2$}\\
  \colhead{(MHz)}&
  \colhead{(Jy)}&
  \colhead{}&
   \colhead{}&
    \colhead{}&
  \colhead{}
  }
  \startdata
81.25 & 0.95 &5318$\pm$1384 & 1.49$\pm$0.15 & 5038$\pm$1180& 2.55$\pm$0.19\\
93.75 & 0.95 &4850$\pm$928 & 1.64$\pm$0.08 & 4658$\pm$615& 2.43$\pm$0.13\\
106.25& 0.8 &4136$\pm$520 & 1.73$\pm$0.07 & 3708$\pm$355& 2.22$\pm$0.13\\
118.75& 0.8 &3664$\pm$441 & 1.89$\pm$0.08 & 3447$\pm$327& 2.17$\pm$0.12\\
131.25& 0.65 &4432$\pm$444 & 1.71$\pm$0.06 & 3594$\pm$276& 2.20$\pm$0.12\\
143.75& 0.65 &3917$\pm$455 & 1.77$\pm$0.06 & 3327$\pm$240& 2.15$\pm$0.16\\
156.25& 0.5 &3747$\pm$613 & 1.83$\pm$0.10 & 3112$\pm$341& 2.10$\pm$0.12\\
168.75& 0.5 &4269$\pm$1167 & 1.75$\pm$0.13 & 3170$\pm$344& 2.18$\pm$0.18
  \enddata
 \end{deluxetable}

\begin{figure}
\begin{center}
\hspace{-4mm}
\includegraphics[width=6cm, angle=-90]{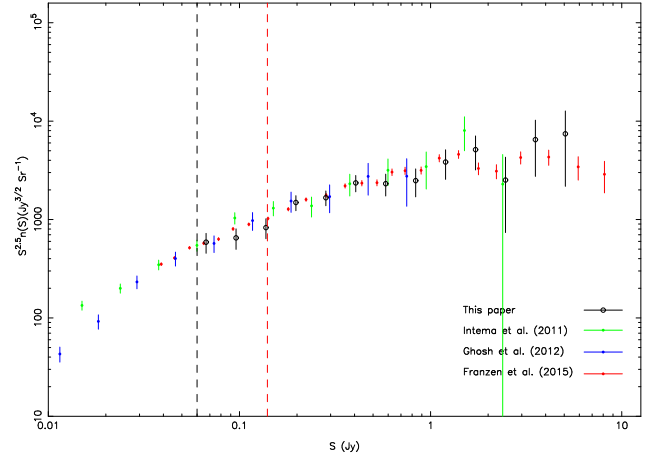}
\vspace{3mm}
\caption{Euclidean normalized differential counts from this paper
at 154 MHz (black points). 
Green and blue points are the GMRT observations at 153 MHz and 150 MHz, 
respectively \citep{Intema11,Ghosh12}. Red points are recent MWA results 
at 154 MHz \citep{Franzen16}. The two vertical dashed lines 
represent $20\%$ (black) and  $50\%$ (red) completeness limits, respectively. }
\vspace{3mm}
\label{fig:154MHz-compare}
\end{center}
\end{figure}

\begin{figure}
\begin{center}
\hspace{-4mm}
\includegraphics[width=6.5cm, angle=-90]{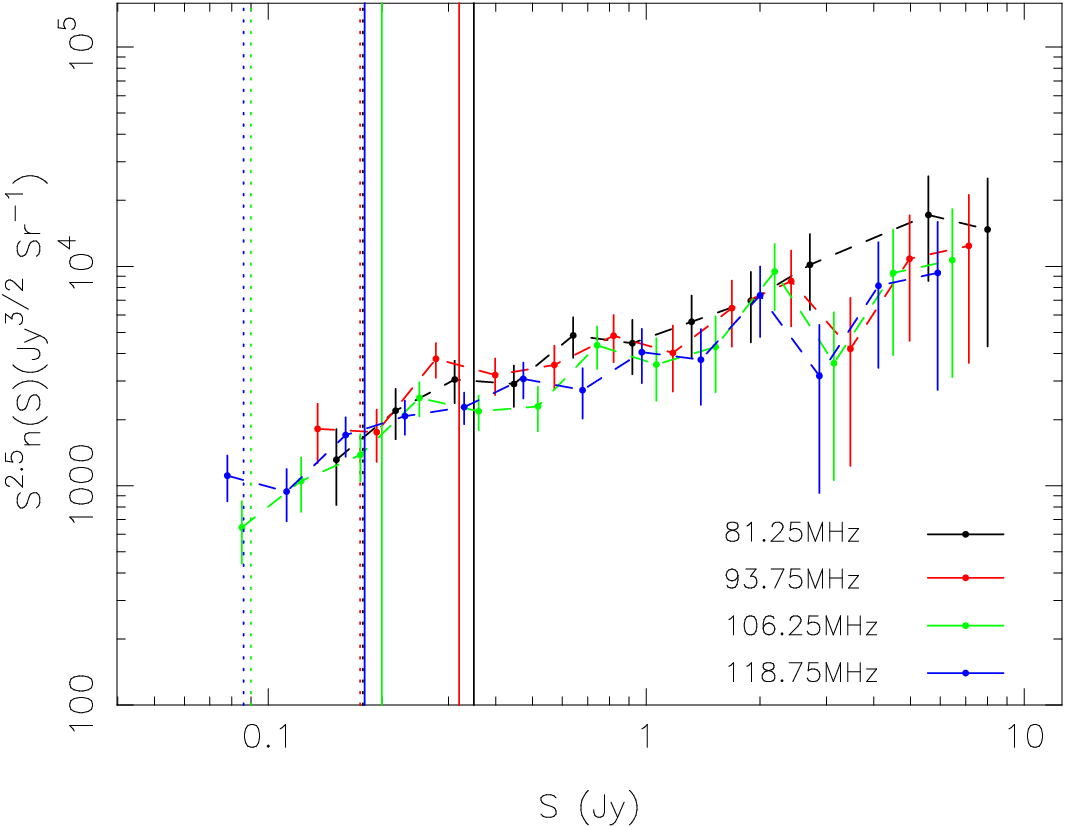}
\vspace{3mm}
\caption{Euclidean normalized differential counts for the lower frequency 
bands (81.25, 93.75, 106.25 and 118.75 MHz).
The vertical lines represent $20\%$ (dotted) and  $50\%$ (solid) completeness
limits.}
\vspace{3mm}
\label{fig:SdNdS-low}
\end{center}
\end{figure}

\begin{figure}
\begin{center}
\hspace{-4mm}
\includegraphics[width=6.5cm, angle=-90]{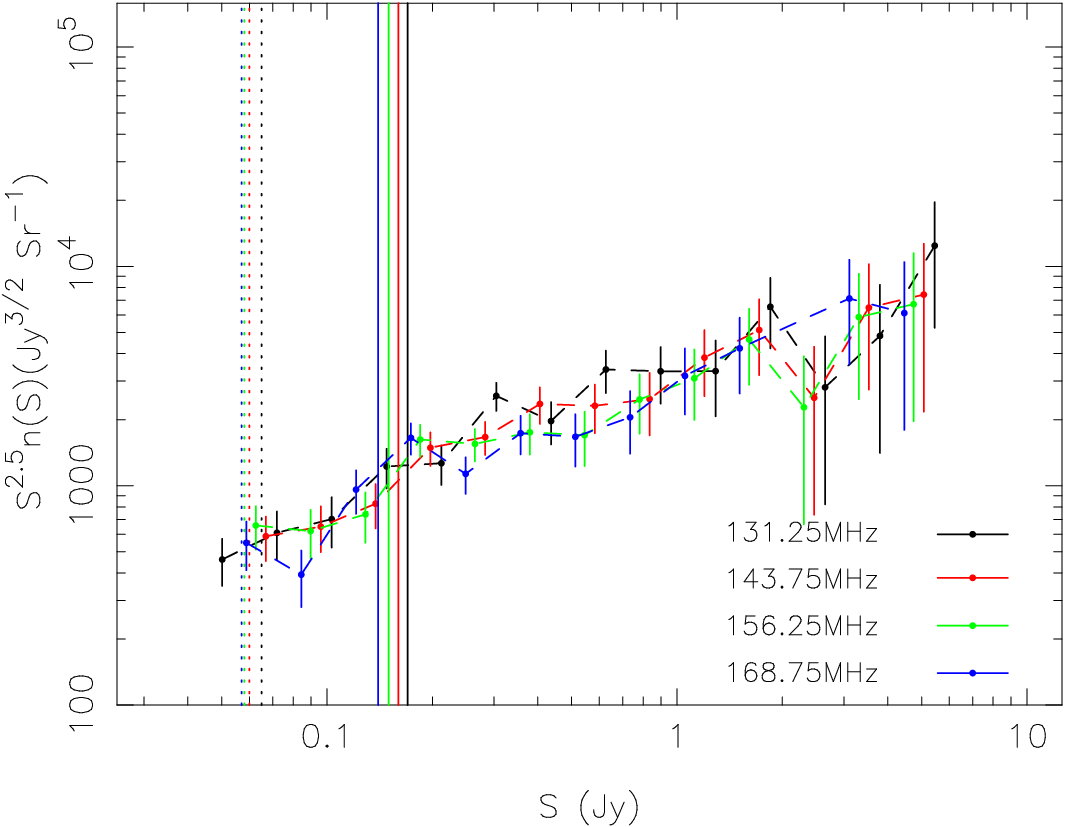}
\vspace{3mm}
\caption{Euclidean normalized differential counts for the higher frequency 
bands (131.25, 143.75, 156.25 and 168.75 MHz).
The vertical lines represent $20\%$ (dotted) and  $50\%$ (solid) completeness
limits.}
\vspace{3mm}
\label{fig:SdNdS-high}
\end{center}
\end{figure}

To compare our results with other deep surveys at low frequencies, 
we derive the Euclidean-normalized differential source counts, 
$S^{5/2}dN/dS$, for which uncertianties are represented by the Poisson 
errors calculated from the number of sources detected in each flux density bin.
We take the source counts from two recent surveys 
at frequencies in the range of 151-154 MHz with GMRT \citep{Intema11,Ghosh12} 
and MWA \citep{Franzen16}. We choose our results in the 156.25 MHz 
band to compare with $S^{5/2}dN/dS$ given at 151-154 MHz with the GMRT 
and MWA by scaling the observing frequencies with a single power-law 
of $S\propto \nu^{-0.8}$ to 154 MHz (Figure \ref{fig:154MHz-compare}). 
Within the error bars, our source counts are in good agreement with these 
recent surveys. 
In Table \ref{nor-counts81-118} and Table \ref{nor-counts131-168} 
we have provided the tabulated normalized differential source counts 
for the lower (81.25, 93.75, 106.25 and 118.75 MHz) and higher 
(131.25, 143.75, 156.25 and 168.75 MHz) bands, respectively, and 
the results are also displayed in Figure \ref{fig:SdNdS-low} and 
Figure \ref{fig:SdNdS-high}, respectively.

\begin{deluxetable*}{lllll}
 \tabletypesize{\scriptsize}
  \tablecaption{Normalized  
differential counts in the lower frequency bands (81.25, 93.75, 106.25 and 
118.75 MHz) from the 21CMA NCP field. $\rm N_{\rm raw}$ is the raw number counts
in each frequency band.
   \label{nor-counts81-118}}
  \tablehead{
  \colhead{$\rm Bin_{\rm start}$}&
  \colhead{$\rm Bin_{\rm end}$}&
  \colhead{$\rm N_{\rm raw}$}&
  \colhead{Normalized counts}&
  \colhead{Completeness}\\
  \colhead{(Jy)}&
  \colhead{(Jy)}&
  \colhead{}&
  \colhead{($\rm Jy^{3/2}sr^{-1}$)}&
 \colhead{correction factor}
  }
  \startdata
81.25 MHz\\
\hline
\hline
0.367& 0.527&    22&  2910$\pm$620 & 1.3\\    
0.527& 0.755&    23& 4841$\pm$1009 & 1.2\\    
0.755& 1.083& 13&  4462$\pm$1237 & 1.1\\   
1.083& 1.554&  10&  5587$\pm$1767  & 1.0\\    
1.554& 2.228&  8&   6964$\pm$2462  & 1.0\\    
2.228& 3.196&  7&   10172$\pm$3844  & 1.0\\    
4.584& 6.575&  4&   17150$\pm$8575  & 1.0\\    
6.575& 9.430&  2&  14730$\pm$10415 & 1.0\\
\hline
\hline
93.75 MHz\\
\hline 
\hline
0.327 & 0.470  & 28 & 3198$\pm$604 & 1.3\\
0.470 & 0.674 &20 & 3553$\pm$794 & 1.2\\
0.674 & 0.966 & 17 & 4830$\pm$1171 & 1.1\\
0.966 & 1.386 & 9 & 4030$\pm$1343 & 1.0\\
1.386 & 1.987 &  9 & 6453$\pm$2151 & 1.0\\
1.987 & 2.850 & 7 & 8567$\pm$3238 & 1.0\\
2.850 & 4.088 & 2 & 4204$\pm$2973 & 1.0\\
4.088 & 5.863 &  3 & 10833$\pm$6254 & 1.0\\
5.863 & 8.410 &  2 & 12405$\pm$8772 & 1.0\\
\hline
\hline
106.25 MHz\\
\hline
\hline
0.207 & 0.296 & 32 & 2513$\pm$444 & 1.9\\
0.296 & 0.425 &  31 & 2183$\pm$392 & 1.0\\
0.425 & 0.609 & 19 & 2299$\pm$527 & 1.0\\
0.609 & 0.874 &  21 & 4364$\pm$952 & 1.0\\
0.874 & 1.254 & 10 & 3570$\pm$1129 & 1.0\\
1.254 & 1.798 &  7 & 4292$\pm$1622 & 1.0\\
1.798 & 2.579 &  9 & 9479$\pm$3160 & 1.0\\
2.579 & 3.699 &  2 & 3618$\pm$2558 & 1.0\\
3.699 & 5.305 & 3 & 9322$\pm$5382 & 1.0\\
5.305 & 7.608 & 2 & 10675$\pm$7459  & 1.0\\
\hline
\hline
118.75 MHz\\
\hline
\hline
0.189 & 0.271 &  32 & 2075$\pm$367 & 1.8\\
0.271 & 0.389 &  37 & 2280$\pm$375 & 1.0\\
0.389 & 0.557 & 29 & 3070$\pm$570 & 1.0\\
0.557 & 0.800 & 15 & 2728$\pm$704 & 1.0\\
0.800 & 1.147 &  13 & 4061$\pm$1126 & 1.0\\
1.147 & 1.645 &  7 & 3756$\pm$1420 & 1.0\\
1.645 & 2.359 &  8 & 7373$\pm$2606 & 1.0\\
2.359 & 3.384 &  2 & 3166$\pm$2239 & 1.0\\
3.384 & 4.853 &  3 & 8158$\pm$4710 & 1.0\\
4.853 & 6.961 &  2 & 9341$\pm$6605 & 1.0
 \enddata
 \end{deluxetable*}
\begin{deluxetable*}{lllll}
 \tabletypesize{\scriptsize}
  \tablecaption{Normalized  differential counts in the higher frequency 
bands (131.25, 143.75, 156.25 and 168.75 MHz) from the 21CMA NCP field.
$\rm N_{\rm raw}$ is the raw number counts in each frequency band.
   \label{nor-counts131-168}}
  \tablehead{
  \colhead{$\rm Bin_{\rm start}$}&
  \colhead{$\rm Bin_{\rm end}$}&
  \colhead{$\rm N_{\rm raw}$}&
  \colhead{Normalized counts}&
\colhead{Completeness}\\
  \colhead{(Jy)}&
  \colhead{(Jy)}&
  \colhead{}&
  \colhead{($\rm Jy^{3/2}sr^{-1}$)}&
 \colhead{correcton factor}
  }
  \startdata
131.25 MHz\\
\hline
\hline
0.174 & 0.250  & 24 & 1266$\pm$258& 1.6\\
0.250 & 0.359 &  47 & 2569$\pm$375& 1.0\\
0.359 & 0.515 &  21 & 1972$\pm$430& 1.0\\
0.515 & 0.738 &  21 & 3387$\pm$739& 1.0\\
0.738 & 1.059 &  12 & 3324$\pm$960& 1.0\\
1.059 & 1.518 &  7 & 3331$\pm$1259& 1.0\\
1.518 & 2.178 &  8 & 6538$\pm$2312& 1.0\\
2.178 & 3.123 &  2 & 2807$\pm$1985& 1.0\\
3.123 & 4.480 & 2 & 4823$\pm$3410& 1.0\\
4.480 & 6.425 &  3 & 12427$\pm$7174& 1.0\\
\hline
\hline
143.75 MHz\\
\hline
\hline
0.162 & 0.233 &  33 & 1489$\pm$259& 1.5\\
0.233 & 0.334 & 34 & 1666$\pm$286& 1.0\\
0.334 & 0.478 & 28 & 2357$\pm$445& 1.0\\
0.478 & 0.686 &  16 & 2313$\pm$578& 1.0\\
0.686 & 0.984 &  10 & 2484$\pm$785& 1.0\\
0.984 & 1.411 &  9  & 3839$\pm$1279& 1.0\\
1.412 & 2.025 &  7 & 5129$\pm$1939& 1.0\\
2.025 & 2.904 &  2 & 2517$\pm$1780& 1.0\\
2.904 & 4.165 &  3 & 6486$\pm$3745& 1.0\\
4.165 & 5.974 &  2 & 7428$\pm$5252& 1.0\\
\hline
\hline
156.25 MHz\\
\hline
\hline
0.152 & 0.218 &  36 & 1624$\pm$271& 1.7\\
0.218 & 0.312 & 35 & 1551$\pm$262& 1.0\\
0.312 & 0.448 &  23 & 1752$\pm$365& 1.0\\
0.448 & 0.642 &  13 & 1701$\pm$472& 1.0\\
0.642 & 0.921 & 11 & 2472$\pm$745& 1.0\\
0.921 & 1.321 &  8 & 3088$\pm$1092& 1.0\\
1.321 & 1.894 &  7 & 4641$\pm$1754& 1.0\\
1.894 & 2.717 &  2 & 2278$\pm$1611& 1.0\\
2.717 & 3.896 &  3 & 5869$\pm$3388& 1.0\\
3.896 & 5.589 &  2 & 6720$\pm$4752& 1.0\\
\hline
\hline
168.75 MHz\\
\hline
\hline
0.142 & 0.205 &  38 & 1653$\pm$268& 1.8\\
0.205 & 0.293 &  28 & 1132$\pm$214& 1.0\\
0.293 & 0.421 &  25 & 1736$\pm$347& 1.0\\
0.421 & 0.604 &  14 & 1670$\pm$446& 1.0\\
0.603 & 0.866 &  10 & 2049$\pm$648& 1.0\\
0.866 & 1.242 &  9 & 3167$\pm$1056& 1.0\\
1.242 & 1.781 &  7 & 4232$\pm$1599& 1.0\\
2.554 & 3.664 &  4 & 7135$\pm$3567& 1.0\\
3.664 & 5.255 &  2 & 6128$\pm$4333& 1.0
  \enddata
 \end{deluxetable*}

\subsection{Noise Analysis}
We now analyze the noise in our interferometric imaging towards the NCP region,
which comprises primarily four contributors: thermal noise, confusion noise,
calibration error and deconvolution error.

The thermal noise for a radio interferometer like 21CMA with an effective 
area $A_{\rm eff}$ and system temperature $T_{\rm {\rm sky}}$ can be estimated 
simply from
\begin{equation}
S_{\rm thermal}=\frac{\sqrt{2}k_{\rm B}T_{\rm sys}}{A_{\rm eff}\eta
\sqrt{\triangle\nu\triangle t}},
\end{equation}
where $k_{\rm B}$ is Boltzmann's constant, $\Delta\nu$,  $\Delta t$ and 
$\eta$ are the frequency bandwidth, observation duration, and 
efficiency factor, respectively. $T_{\rm sys}$ is composed of 
the telescope temperature (50K for the 21CMA) and the sky (the Milky Way) 
temperature described approximately by $60(\nu/300{\rm MHz})^{-2.55}$ 
(e.g. \citealt{Chapman12}). For the observation made in this work, 
we have only used 559 baselines among 780 along the E-W arm 
after the shorter and longer baselines are taken out. This reduces 
the total effective area to $A_{\rm eff}=\sqrt{559} a_{\rm eff}$, in which 
$a_{eff}\approx216 m^2$ is the effective area of a single pod which 
remains roughly constant over our observing frequency range in terms 
of our antenna design. We also adopt the current efficiency of 
$\eta=50\%$ for 21CMA to provide the numerical estimation. 

Confusion noise $\sigma_c$ from the position uncertainties of unresolved, 
faint radio sources below a flux threshold $S_{\rm lim}$ with finite synthesis 
beam $\Omega_{\rm b}$ can be estimated through (e.g. \citealt{Scheuer57}; 
\citealt{Condon74}) 
\begin{equation}
\sigma_c^2=\Omega_{\rm b}\int_0^{S_{lim}}\frac{dN}{dS}S^2dS.
\end{equation}
In principle, we could take a simple extrapolation of our current 
differential source counts $dN/dS$ to lower fluxes beyond $S_{\rm lim}$ 
by considering a lower completeness of our catalog especially at lower 
frequencies. Here we would rather choose the deeper GMRT observations 
at 153 MHz \citep{Intema11} and extrapolate their differential number 
counts to $S_{\rm lim}=0.1$ mJy, two orders of magnitude fainter than their 
flux limit. Yet, uncertainty in this extrapolation to very low flux remains 
unclear. Moreover, it seems that the steep power law for the number count 
may flatten towards lower flux below a few mJy 
(see \citealt{Franzen16} for a recent compilation). 

\begin{figure}
\begin{center}
\hspace{-4mm}
\includegraphics[width=8cm]{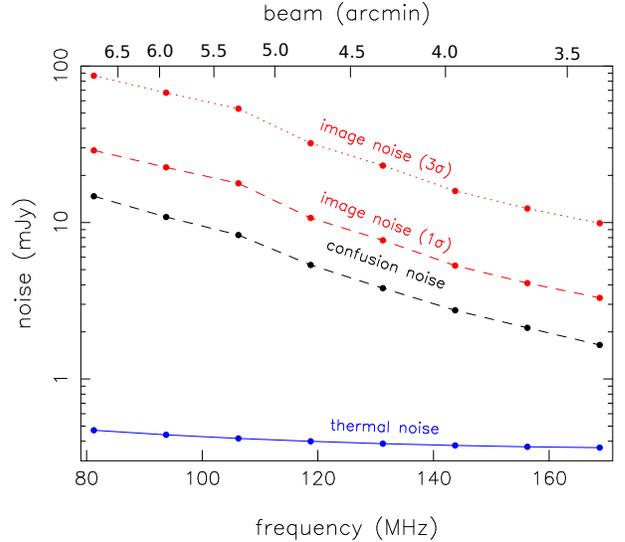}
\vspace{3mm}
\caption{Sensitivity and noise as a function of observing frequencies
for a bandwidth of 12.5 MHz. The blue and solid line denotes the expected 
thermal noise of the 21CMA for a 12 hour integration. Red points, together 
with dashed and dotted lines, are the $1\sigma$ and $3\sigma$ noise levels, 
respectively, measured at eight sub-bands over the NCP field. The black 
dashed line represents the theoretically estimated confusion limit due to 
the unresolved fainter sources below $S_{\rm lim}$, for which we have 
taken $S_{\rm lim}=3\sigma$. }
\vspace{3mm}
\label{fig:confusion}
\end{center}
\end{figure}

Figure \ref{fig:confusion} plots the theoretically predicted thermal 
noise and confusion noise, together with the image noise measured across 
all eight bands. We have set  $S_{\rm lim}=3\sigma$ in the estimate of confusion 
limit in each of the subbands. Thermal noise for 12 hour observations and a bandwidth 
of 12.5 MHz has already become smaller than the image noise, indicating that 
the key factor influencing the sensitivity of low-frequency interferometric 
imaging is no longer the system noise. Indeed, the confusion limit arising 
mainly from the unresolved faint sources especially at lower frequencies 
has actually made a significant contribution to the noise level. Unlike 
the thermal noise which decreases as $t^{-1/2}$, the confusion noise is 
very insensitive to observing time $t$. Instead, the upper limit of integral 
in Eq.(6), $S_{\rm lim}$, now determines the variation trend in confusion 
limit. One of our primary tasks in low-frequency interferometric imaging 
is to suppress the low flux limit $S_{\rm lim}$ to allow more fainter sources to 
be identified and resolved.

In fact, $S_{\rm lim}$ is controlled by both calibration error and deconvolution 
error, in addition to telescope resolution. Here the current imaging algorithm 
has not taken ionospheric effects into account, though self-calibration may 
partially correct the effect if a timestep is chosen to be within 
a few minutes or only short baselines are used. Performing calibration on a timescale of a 
few minutes turns out to be very time consuming since the correction should 
also be made at each frequency channel. Moreover, the grating lobes due to 
numerous redundant baselines of the 21CMA generate very prominent, equally 
spaced rings round bright sources such as B004713+891245 and 3C061.1. Complete 
removals of these ring-like structures in deconvolution processing is still 
difficult at present. First, an accurate sky model containing both bright point 
and extended sources should be constructed in the calibration processing. 
Significant uncertainties arise from the inaccurate models of (marginally) 
resolved extended sources, in which surface brightness, spectral index and 
position of each component of the sources should be determined  by other radio 
observations or left as free parameters in the sky model. The brightest radio 
galaxy, 3C061.1, in the NCP field is a typical example of this kind, and its 
residuals after deconvolution are very prominent. Actually, 3C061.1 is the 
dominant source of both calibration and deconvolution errors for our imaging 
towards the NCP region. Note that calibration error is also present even if 
a more sophisticated model for 3C061.1 is adopted for LOFAR data 
\citep{Yatawatta13}. Second, the deconvolution should be made 
over a sky map with the field-of-view much larger than the primary 
beam so that the leakage of sidelobes from off-field strong radio 
sources can be corrected for. For the 21CMA field, the two brightest 
radio sources, Cas A and Cygnus A, are visible on the sky map of 
$60^{\circ}\times60^{\circ}$ (\citealt{Huang16}), and their grating lobes 
have already entered into the central NCP field at certain frequencies. 
At present the imaging has not included and corrected the sidelobes from 
these two sources, although they have already contributed to the deconvolution 
noise.

In summary, it appears that with time integration the thermal noise 
dominated by the Milky Way can be well controlled to below the confusion 
noise in current low-frequency interferometric imaging. However, to beat 
down the confusion noise, one needs to deal with not only the unresolved, 
faint radio sources, but also the calibration and deconvolution errors. 
In particular, calibration errors from both ionospheric effects and bright, 
extended sources are difficult to handle at present. These errors 
can propagate to the deconvolution processing and affect the lower threshold 
of identifying the faint radio sources through $S_{\rm lim}$. As a result, 
the conventional confusion limit, which is determined by unresolved sources 
and synthesized beam, is actually affected by the extension of accuracy and 
reliability of calibration and deconvolution. In fact, the major source of 
errors in our current imaging arises from the calibration and deconvolution 
processing. The former is related to both ionospheric effects and modeling 
of extended, bright radio sources in the field, and the latter is involved 
with the removals of grating lobes and sidelobes of strong radio sources, 
both within and outside of the field.    

\section{Discussion and Conclusions}

We have analyzed 12 hours of data taken from the 21CMA observations 
centered on the NCP.  To reduce the complexity of interferometric image 
processing and the influence of ionospheric perturbation, we have used 
all 40 pods along the east-west arm but restricted our analysis to 
within a maximum baseline of 1500 m. We have calibrated the gain and 
primary beam of our telescope using the bright radio sources in the NCP 
field. Following conventional self-calibration and deconvolution methods, 
we have detected a total of 624 radio sources over the central field 
within $3^{\circ}$ in frequency range of 75 - 175 MHz and the outer 
annulus of $3^{\circ}-5^{\circ}$ in the 75 - 125 MHz bands. By performing a 
Monte-Carlo simulation we have estimated a completeness of
$50\%$ at $S\sim0.2$ Jy.  We have compared our source counts with 
the deep low-frequency observations made recently with GMRT and MWA, 
and the completeness-corrected source counts show a good agreement with 
these recent surveys at the corresponding frequency bands.

While we are able to detect the fainter sources down to 10 mJy, the
detection fraction has dropped to $\sim10\%$, making a ultra deep 
observation rather difficult and even impossible with current imaging 
algorithms. This is primarily caused by errors in sidelobes, calibration,
and deconvolution as well as the confusion limit of the array.  Spacings 
between the 21CMA pods are the integral multiples of 20 m. The original 
design of such configuration is for the purpose of redundant calibration 
and statistical measurement of the EoR power spectrum 
(e.g. \citealt{Noordam82,Tegmark10}). However, this layout also generates 
the grating lobes evident in our images (\citealt{Bracewell73,Amy90}). 
These equally spaced rings around the bright sources in the field have 
their brightness almost comparable to the central sources, and therefore 
should be accurately removed in the deconvolution processing. Otherwise,
their residuals may be the major sources of contaminations for our
imaging. Indeed, our image quality is largely limited by the presence
of imperfectly subtracted grating lobes of the brightest source in the 
NCP field, 3C061.1. Although this arises partially from the inaccurate 
calibration, the main reason is  the poor modeling of 3C061.1 in our  
self-calibration processing. 3C061.1 is marginally resolved with 21CMA,
and its modeling should take the structured components with different 
surface brightness and spectral index into account 
(e.g. \citealt{Yatawatta13}). Our next step is to improve our sky model 
by using a sophisticated model for 3C061.1 and even for very bright, 
far field sources such as Cas A and Cygnus A. Finally, the theoretically 
expected thermal noise ($\sim0.4$ mJy) is well below the confusion limit due 
to the unresolved, fainter sources, indicating that the dominated 
factor of influencing the sensitivity for low-frequency interferometric 
imaging is no longer the system noise but the cosmic sources in the sky. 

Several methods to improve our imaging algorithm including an accurate 
calibration (e.g. sky modeling and redundant calibration), application of 
the w-term correction, and ``peeling'' sources \citep{Noordam04} are currently 
under investigation, with our ultimate goal towards detecting the EoR power 
spectrum. Our experience with the 21CMA operation and data analysis may 
provide a useful guide to the design of next generation low-frequency radio 
array such as the SKA. Indeed, employment of numerous redundant baselines 
helps to improve the precision of calibration. However, it brings about 
very prominent grating lobes from bright sources both in-beam and in the far-field. 
This not only requires a more careful and accurate calibration of visibilities
but also adds an extra difficulty to deconvolution processing. In particular, 
the grating lobes of the bright, structured sources like giant radio galaxies 
with jets are hard to model and thus subtract. In this sense, both random 
antenna element and station layouts should preferentially be chosen for 
low-frequency radio interferometers to suppress the grating lobes and 
sidelobes of bright sources. We note that an extensive study on this topic 
has recently been carried out by \citet{Mort16}. 

To summarize, with current layout of the 21CMA and conventional imaging 
algorithm, we have reached a sensitivity of a few mJy and a dynamical 
range of $10^4$. We may have to improve our imaging quality by an order 
of magnitude in order to see statistically the EoR signature.

\section{Acknowledgments}

We thank Judd Bownman for useful discussions and Heinz Andernach for constructive comments.
We gratefully acknowledge the constructive suggestions by an 
anonymous referee that greatly improved the presentation of this work. Technical 
support was provided by the 21CMA collaboration. This work was partially 
supported by the National Science Foundation of China under Grant No. 11433002. 
QZ and MJ-H are supported in this work through a Marsden Fund grant to MJ-H in 
New Zealand. The 21CMA is jointly operated and administrated by National Astronomical
Observatories of China and Center for Astronomical Mega-Science, Chinese 
Academy of Sciences. This research has made use of the NASA/IPAC Extragalactic
Database (NED), which is operated by the Jet Propulsion Laboratory, California
Institute of Technology, under contract with the National Aeronautics and
Space Administration.


\newpage

\begin{thebibliography}{}
\bibitem[\protect\citeauthoryear{Amy \& Large}{1990}]{Amy90}
Amy, S. W., \& Large, M. I. 1990, Astronomical Society of Australia, 8, 308
\bibitem[\protect\citeauthoryear{Barkana \& Loeb}{2001}]{Barkana01}
Barkana, R., \& Loeb, A. 2011, PhR, 349, 125
\bibitem[\protect\citeauthoryear{Bernardi et al.}{2010}]{Bernardi10}
Bernardi, G., de Bruyn, A. G., Harker, G., et al. 2010, \aap,522, A67
\bibitem[\protect\citeauthoryear{Bowman et al.}{2007}]{Bowman07}
Bowman, J.~D., Barnes, D. ~G., Briggs, F. ~H., et al. 2007, \aj, 133, 1505
\bibitem[\protect\citeauthoryear{Bowman et al.}{2013}]{Bowman13}
Bowman, J.~D., Cairns, I., Kaplan, D., et al. 2013, PASA, 30, 31
\bibitem[\protect\citeauthoryear{Bracewell \& Thompson}{1973}]{Bracewell73}
Bracewell, R. N., \& Thompson, A. R. 1973, \apj, 182, 77
\bibitem[\protect\citeauthoryear{Chapman et al.}{2012}]{Chapman12}
Chapman, E., Abdalla, F. B., Harker, G., et al. 2012, \mnras, 423, 2518
\bibitem[\protect\citeauthoryear{Cohen}{2007}]{Cohen07}
Cohen, A. S., Lane, W. M., Cotton, W. D., et al. 2007, \aj, 134, 1245
\bibitem[\protect\citeauthoryear{Condon}{1974}]{Condon74}
Condon, J. J. 1974, \apj, 188, 279
\bibitem[\protect\citeauthoryear{Franzen et al.}{2016}]{Franzen16}
Franzen, T. M. O., Jackson, C. A., Offringa, A. R., et al. \mnras, 459, 3314
\bibitem[\protect\citeauthoryear{Furlanetto, Oh \& Briggs}{2006}]{Furlanetto06}
Furlanetto, S. R., Oh, S. P., \& Briggs, F. H. 2006, PhR, 433, 181
\bibitem[\protect\citeauthoryear{Ghosh et al.}{2011}]{Ghosh11}
Ghosh, A., Bharadwaj, S., Ali, S. ~S., \& Chengalur, J. ~N. 2011, \mnras, 418, 2584
\bibitem[\protect\citeauthoryear{Ghosh et al.}{2012}]{Ghosh12}
Ghosh, A., Prasad, J., Bharadwaj, S., Ali, S. ~S., \& Chengalur, J. ~N. 2012, 
\mnras, 426, 3295
\bibitem[\protect\citeauthoryear{Hales, Baldwin \& Warner}{1988}]{Hales88}
Hales, S. E., Baldwin, J. E., \& Warner, P. J. 1988, \mnras, 234, 919
\bibitem[\protect\citeauthoryear{Hales et al.}{2007}]{Hales07}
Hales, S. E. G., Riley, J. M., Waldram, E. M., Warner, P. J., 
Baldwin, J. E. 2007, \mnras, 382, 1639
\bibitem[\protect\citeauthoryear{Heald et al.}{2015}]{Heald15}
Heald, G. ~H., Pizzo, R. ~F., Orr{\'u}, E., et al. 2015, arXiv:1509.01257
\bibitem[\protect\citeauthoryear{Heywood et al.}{2013}]{Heywood13}
Heywood, I., Jarvis, M. J., \& Condon, J. J. 2013, \mnras, 432, 2625
\bibitem[\protect\citeauthoryear{H{\"o}gbom}{1974}]{hog74}
H{\"o}gbom, J.~A. 1974, \aaps, 15, 417
\bibitem[\protect\citeauthoryear{Holder}{2012}]{Holder12}
Holder, J. 2012, Astroparticle Physics, 39, 61
\bibitem[\protect\citeauthoryear{Huang et al.}{2016}]{Huang16}
Huang,Y., Wu, X. -P., Zheng, Q., Gu, J. -H., \& Xu, H. 2016, RAA, 16, 16
\bibitem[\protect\citeauthoryear{Hurley-Walker et al.}{2014}]{NHW14}
Hurley-Walker, N., Morgan, J., Wayth, R. ~B., et al. 2014, PASA, 31, 45
\bibitem[\protect\citeauthoryear{Intema et al.}{2011}]{Intema11}
Intema, H. ~T., van Weeren, R. ~J., R{\"o}ttgering, H. ~J. ~A., \& Lal, D. ~V. 
2011, \aap, 535, A38
\bibitem[\protect\citeauthoryear{Intema et al.}{2016}]{Intema16}
Intema, H. ~T., Jagannathan, P., Mooley, K. P., Frail, D. A.
2016, arXiv:1603.04368
\bibitem[\protect\citeauthoryear{Jacobs et al.}{2011}]{Jacobs11}
Jacobs, D. ~C., Aguirre, J. ~E., Parsons, A. ~R.,et al. 2011, \apjl, 734, L34.
\bibitem[\protect\citeauthoryear{Jeli{\'c} et al.}{2008}]{Jelic08}
Jeli{\'c}, V., Zaroubi, S., Labropoulos, P., et al. 2008, \mnras, 389, 1319
\bibitem[\protect\citeauthoryear{Jeli{\'c} et al.}{2014}]{Jelic14}
Jeli{\'c}, V., de Bruyn, A. ~G., Mevius, M., et al. 2014, \aap, 568, A101
\bibitem[\protect\citeauthoryear{Kazemi et al.}{2011}]{Kazemi11}
Kazemi, S., Yatawatta, S., Zaroubi, S.,  et al. 2011, \mnras, 414, 1656
\bibitem[\protect\citeauthoryear{Lawrence et al.}{1996}]{Lawrence96}
Lawrence, C. R., Zucker, J. R., Readhead, A. C. S., et al., 1996, ApJS, 107, 541
\bibitem[\protect\citeauthoryear{Massaro et al.}{2014}]{Massaro14}
Massaro, F., Giroletti, M., D'Abrusco, R., et al. 2014, \apjs, 213, 3 
\bibitem[\protect\citeauthoryear{Moore et al.}{2013}]{Moore13}
Moore, D. F., Aguirre, J. E., Parsons, A. R., Jacobs, D. C., \& Pober, J. C. 
2013, \apj, 769, 154
\bibitem[\protect\citeauthoryear{Mort et al.}{2016}]{Mort16}
Mort, B., Dulwich, F., Razavi-Ghods, N., de Lera Acedo, E., \& Grainge, K. 
2016, arXiv:1602.01805
\bibitem[\protect\citeauthoryear{Noordam \& de Bruyn}{1982}]{Noordam82}
Noordam, J. E., \& de Bruyn A. G. 1982, \nat, 299, 597
\bibitem[\protect\citeauthoryear{Noordam}{2004}]{Noordam04}
Noordam, J. E. 2004, Ground-based Telescopes. Edited by Oschmann, Jacobus M., 
Jr. Proceedings of the SPIE, 5489, 817 
\bibitem[\protect\citeauthoryear{Offringa et al.}{2015}]{Offringa15}
Offringa, A. R., Wayth, R. B.,Hurley-Walker, N., et al. 2015, PASA, 32, 8
\bibitem[\protect\citeauthoryear{Offringa et al.}{2016}]{Offringa16}
Offringa, A. R., Trott, C. M., Hurley-Walker, N., et al. 2016, \mnras, 459, 3314
\bibitem[\protect\citeauthoryear{Paciga et al.}{2013}]{Paciga13}
Paciga, G., Albert, J. G., Bandura, K., et al. 2013, \mnras, 433,639
\bibitem[\protect\citeauthoryear{Pritchard \& Loeb}{2010}]{Pritchard10}
Pritchard, J. R., \& Loeb, A. 2010, PhRvD, 82, 3006
\bibitem[\protect\citeauthoryear{Rees}{1990}]{Rees90}
Rees, N. 1990, \mnras, 244, 233
\bibitem[\protect\citeauthoryear{Scheuer}{1957}]{Scheuer57}
Scheuer, P. A. G. 1957, PCPS, 53, 764
\bibitem[\protect\citeauthoryear{Slee}{1995}]{Slee95}
Slee, O. B. 1995, Australian Journal of Physics, 48, 143
\bibitem[\protect\citeauthoryear{Taylor et al.}{2012}]{Taylor12}
Taylor, G. B., Ellingson, S. W., Kassim, N. E., et al. 2012, 
Journal of Astronomical Instrumentation, 1, 50004
\bibitem[\protect\citeauthoryear{Tegmark \& Zaldarriaga}{2010}]{Tegmark10}
Tegmark, M., \& Zaldarriaga, M. 2010,PhRvD, 82, 103501
\bibitem[\protect\citeauthoryear{Thyagarajan et al.}{2015}]{Thyagarajan15}
Thyagarajan, N., Jacobs, D. C., Bowman, J. D., et al. 2015, \apj, 804, 14
\bibitem[\protect\citeauthoryear{Tingay et al.}{2013}]{Tingay13}
Tingay, S. ~J., Geoke, R., Bowman, J. ~D., et al. 2013, PASA, 30, 7
\bibitem[\protect\citeauthoryear{van Haarlem et al.}{2013}]{vH13}
van Haarlem, M. ~P, Wise, M. ~W., Gunst, A. ~W., et al. 2013, \aap, 556, A2
\bibitem[\protect\citeauthoryear{van Weeren}{2014}]{vanWeeren14}
van Weeren R. J., Williams, W. ~L., Tasse, C.,et al. 2014, \apj, 793, 82
\bibitem[\protect\citeauthoryear{Wayth et al.}{2015}]{Wayth15}
Wayth, R. B., Lenc, E., Bell, M. ~E., et al. 2015, PASA, 32, 25
\bibitem[\protect\citeauthoryear{Whiting}{2011}]{Whiting01}
Whiting, M. T. 2011, \mnras, 421, 3242
\bibitem[\protect\citeauthoryear{Williams, Intema \& R{\"o}ttgering}{2013}]
{Williams13}Williams, W. ~L., Intema, H. ~T., \& R{\"o}ttgering, H. ~J. ~A. 
2013, \aap, 549, A55
\bibitem[\protect\citeauthoryear{Yatawatta et al.}{2009}]{Yatawatta09}
Yatawatta, S., Zaroubi, S., de Bruyn, G., Koopmans, L., Noordam, J. 2009, 
arXiv:0810.5751
\bibitem[\protect\citeauthoryear{Yatawatta et al.}{2013}]{Yatawatta13}
Yatawatta, S., de Bruyn, A. G., Brentjens, M. A.,et al. 2013, A\&A, 550, A136




\end{thebibliography}
\end{document}